\documentclass[11pt,a4paper]{article}
\usepackage{jheppub}
\usepackage{mathtools}
\usepackage{amssymb}
\usepackage{amsfonts}
\usepackage{latexsym}
\usepackage{cancel}
\usepackage{pdfpages}
\usepackage{graphicx,color}

\def\al{\alpha}

\def\lag{\mathcal{L}}
\def\covD{\nabla}
\def\pd{\partial}
\def\nn{\nonumber}
\def\mc{\mathcal}
\def\ve{\varepsilon}

\def\be{\begin{equation}}
\def\te{\end{equation}}
\def\bse{\begin{subequations}}
\def\ese{\end{subequations}}
\def\ba{\begin{align}}
\def\ha{\hat{A}}
\def\th{\theta}

\newcommand{\bee}{\begin{equation}}
\newcommand{\een}{\end{equation}}

\def\bea{\begin{eqnarray}}
\def\eea{\end{eqnarray}}
\def\ba{\begin{eqnarray}}
\def\ea{\end{eqnarray}}
\def\p{\partial}
\def\2{\sqrt{2}}

\def\mc{\mathcal}

\def\be{\begin{equation}}
\def\ee{\end{equation}}
\def\bea{\begin{eqnarray}}
\def\eea{\end{eqnarray}}

\def\th{\theta}

\def\la{\lambda}
\usepackage{color}

\def\al{\alpha}

\def\lag{\mathcal{L}}
\def\covD{\nabla}
\def\pd{\partial}
\def\nn{\nonumber}
\def\mc{\mathcal}
\def\ve{\varepsilon}

\def\be{\begin{equation}}
\def\te{\end{equation}}
\def\bse{\begin{subequations}}
\def\ese{\end{subequations}}
\def\ba{\begin{align}}

\title{Galileon Higgs Vortices }

\author[a]{Javier Chagoya,\,}
\affiliation[a]{Departmento de F\'isica, Universidad de Guanajuato, DCI, Campus Le\'on, C.P. 37150,
Le\'on, Guanajuato, M\'exico. }
\emailAdd{jfchagoya@fisica.ugto.mx}
\author[b]{Gianmassimo Tasinato}
\affiliation[b]{Department of Physics, Swansea University, Swansea, SA2 8PP, U.K.}
\emailAdd{g.tasinato@swansea.ac.uk}
\abstract{Vortex solutions are topologically stable field configurations that
can play an important role in condensed matter, field theory, and cosmology.  
We investigate vortex configuration in a 2+1 dimensional Abelian Higgs theory supplemented
by higher order derivative self-interactions, related
with Galileons.
 Our vortex solutions have
 features that make them qualitatively different from well-known
 Abrikosov--Nielsen-Olesen configurations, since the derivative interactions 
 turn on  gauge invariant field profiles that 
 break axial symmetry.  By promoting the system   to a 
3+1 dimensional string configuration, we study  its gravitational backreaction.
Our  results are all derived within a specific, analytically manageable  system, 
  and  
 might offer indications  for  understanding 
 Galileonic interactions and screening mechanisms 
 around configurations that are not spherically symmetric, but only at most cylindrically 
 symmetric.
}

\begin{document}
\maketitle

\flushbottom

\section{Introduction}

Exact solutions to  classical equations of motion 
 offer glimpses on the non-perturbative structure
of  
 a given field theory,
and can have important physical applications. 
  Vortex
solutions (Abrikosov \cite{abrvortex}, Nielsen-Olesen \cite{Nielsen:1973cs})   are a good example since they play an essential role in 
the physics  of superfluidity and superconductivity (see for example \cite{Donnelly}), and might exist 
  in our 3+1 dimensional universe  in the form of cosmic strings (see e.g. \cite{Vilenkin:2000jqa}). In a covariant Abelian Higgs  theory, Nielsen-Olesen (NO) vortex solutions are axially symmetric field configurations of finite energy, characterised by  couplings to the Higgs scalar and the electromagnetic vector field. 
They support a non-vanishing magnetic flux, and  spontaneously break the Abelian gauge symmetry out of the vortex core.  See e.g. \cite{Weinberg:2012pjx, Dunne:1998qy} for excellent textbook discussions on topological solutions in field theories. 
  
  In this paper,  we ask what happens when the standard Abelian Higgs action is modified by adding higher order, gauge invariant  derivative  self-interactions for the Higgs field.  For this aim, we
 focus on  exploring vortex configurations in a Higgs model in $2+1$ dimensions, supplemented by   derivative Higgs self-interactions related with Galileons \cite{Hull:2014bga}. 
  
  Our motivations are the following:

  \begin{itemize}

  \item[$\blacktriangleright$] In the context of scalar tensor theories of gravity, or massive gravity, theories
    with Galileonic  symmetries \cite{Nicolis:2008in} have received much attention. They enjoy powerful non-renormalization
     theorems \cite{Nicolis:2004qq,Nicolis:2008in}, and offer  good control  of strongly coupled regimes 
    that realise a 
    Vainshtein mechanisms to screen  scalar fifth forces  and to raise
    the effective cut-off of the theory (see e.g. \cite{Brax:2013ida, Babichev:2013usa} for reviews) .
         Much 
     of the analytic studies have focussed on spherically symmetric configurations, since the relevant field 
     equations become algebraic \cite{Babichev:2012re, Sbisa:2012zk, Koyama:2013paa}. On the other hand, 
      it is interesting and important to ask what happens to the 
     Vainshtein mechanism for less symmetric
     cases, as for cylindrical configurations (see e.g. \cite{Bloomfield:2014zfa} for a preliminary study on this respect that neglects gravity backreaction, { or the slowly
     rotating solutions discussed in  \cite{Chagoya:2014fza}}). 
      The theory we consider is simpler than systems involving gravity, so it allows us  to  address 
       analytically  the problem of finding  configurations in axially symmetric set-ups.
Some of the relevant equations of motion are   algebraic, 
 making  the analysis particularly straightforward.  At the same time, the action is sufficiently 
 non-linear to lead to   new properties that are absent for the NO vortex, and that              might be shared with systems  where gravity is important.  
       
       \item[$\blacktriangleright$]  For Galileon symmetric theories, an analogue of Derrick theorem applies,  preventing the existence of  vortex configurations of finite energy in systems     with 
scalar derivative self-interactions only 
\cite{Endlich:2010zj} (see also \cite{Padilla:2010ir}).  Recall that 
 in the standard case of a theory of  scalar fields with non-derivative interactions,  the classical Derrick theorem  \cite{Derrick:1964ww}  states that no finite energy vortex
 solutions exist.  This conclusion can be circumvented by
 adding more structure to the theory, for example  gauging the system by introducing vector fields with  appropriate couplings and asymptotic behaviour.  In this work, we  are interested in  a theory 
  characterized by  a   Mexican hat Higgs potential;
 additionally, it   contains higher order Higgs derivative self-interactions related
 with Galileons, and being gauged it 
couples the Higgs to a vector field \cite{Hull:2014bga}. 
   In such a framework, it is natural to ask whether the new Higgs
   derivative interactions qualitatively modify the structure of NO vortex, leading to novel  effects that is worth exploring.
      


\end{itemize}

\bigskip
Starting from these motivations, we add higher order,  derivative self-interactions to the standard Abelian
Higgs Lagrangian. Such interactions have been first introduced in \cite{Hull:2014bga} for providing a Higgs
mechanism to spontaneously break the gauge symmetry through  vector Galileon interactions \cite{Tasinato:2014eka,Heisenberg:2014rta}.  They are ghost free, and in a suitable high energy limit they enjoy Galilean
symmetries that might protect their structure. We dub this system Galileon Higgs model \footnote{See also e.g. \cite{Kamada:2010qe} for   generalizations of the standard
 Higgs model by means of derivative interactions, in the context of inflationary model building.}. 

Within this framework, we are able to determine vortex solutions characterised by topologically conserved
winding numbers. They have features that make them qualitatively different from the NO vortex.  
The derivative  non-linear interactions necessarily turn on  a new gauge invariant  radial vector component. 
This radial component   violates a reflection symmetry around one
of the axis of the Cartesian coordinates,  and leads  to configurations
that   additionally  break the axial symmetry of the configuration. Interestingly, some of the equations of motion 
reduce to quadratic algebraic equations, making our analysis particularly straightforward.



\begin{figure}[!htb!]
\begin{center}
\includegraphics[width=0.37\textwidth]{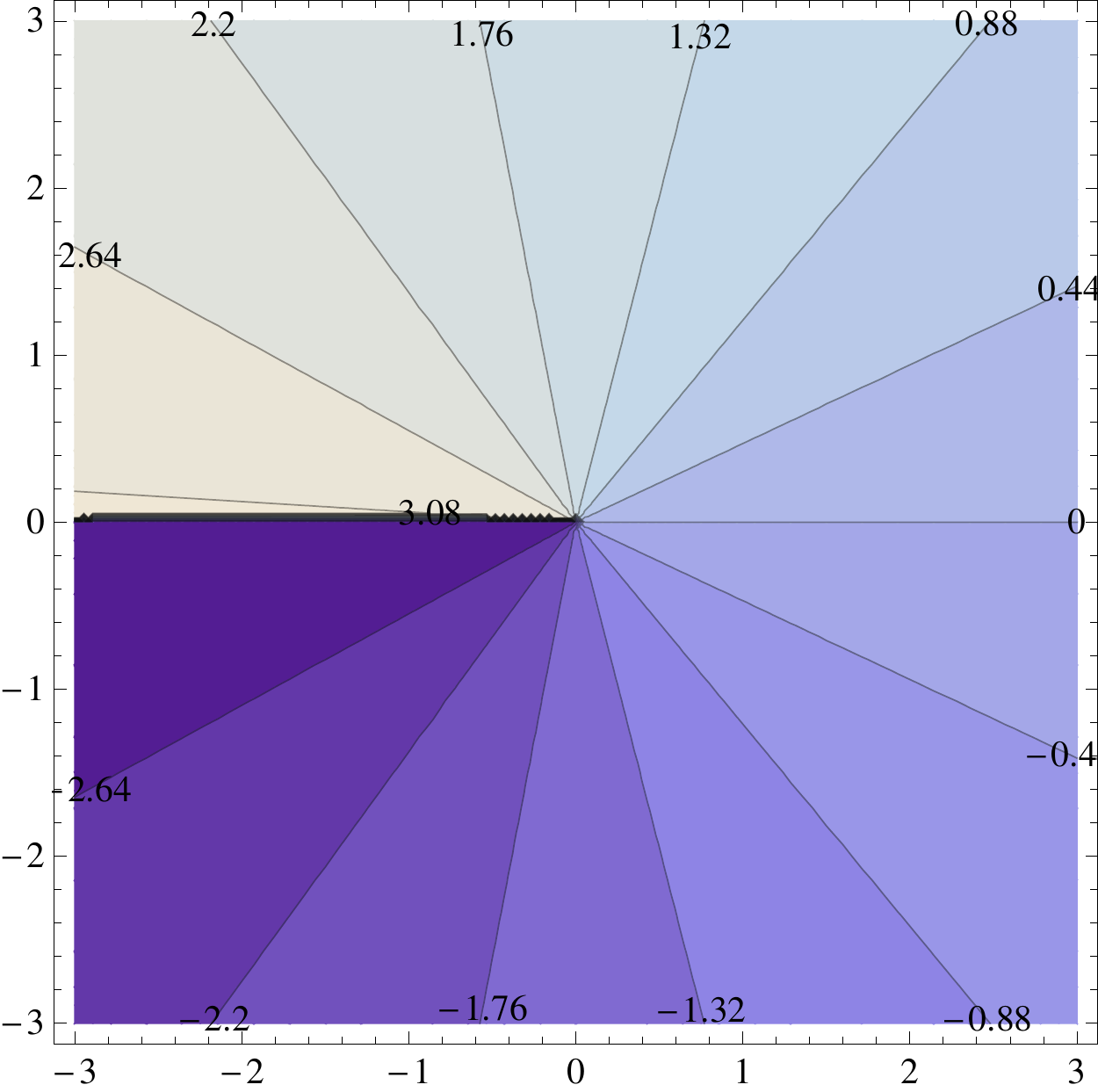}  \ \ \includegraphics[width=0.37\textwidth]{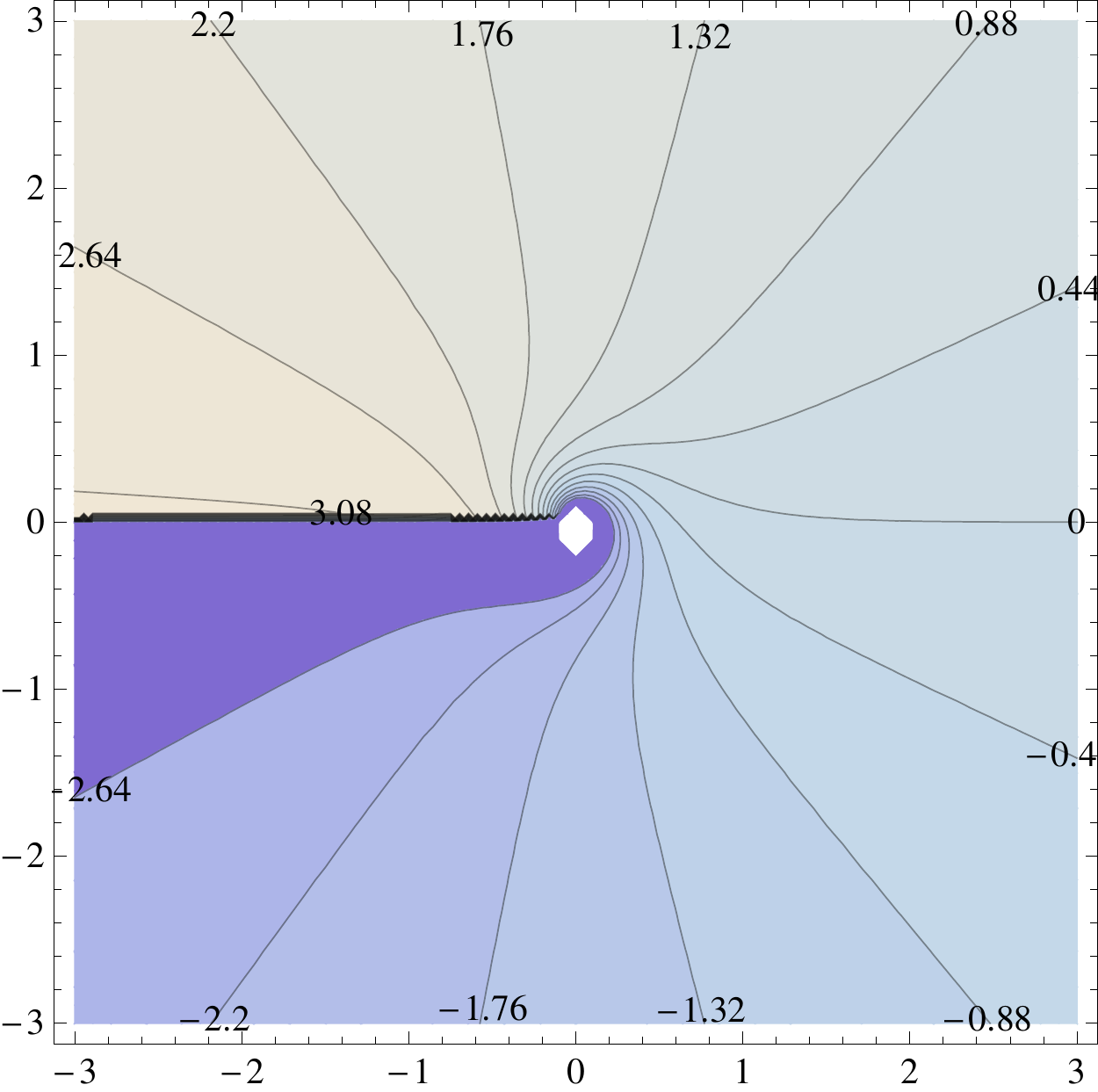} \caption{{\it A representative example of the
 gradient of the  Higgs phase for the Nielsen-Olesen vortex solution (left) and an approximate 
solution for the Galileonic vortex (right), plotted in the $x$-$y$ plane. The lines
 correspond to constant values for the Higgs phase.}}\label{phase2f}
\end{center}
\end{figure}

Figure 
\ref{phase2f} provides 
a graphical  anticipation of our results: 
    we show the gradient of the phase of the scalar field for the NO
solution (left) and our Galileon Higgs vortex configuration (right) in a plane with Cartesian coordinates
$(x_1,\,x_2)$. The violation of a reflection symmetry $x_1\to-x_1$ is manifest, and  axial  symmetry is broken since
surfaces of constant scalar phase are not invariant under rotation of the central axis. { Although in Fig \ref{phase2f} we focus on the 
Higgs phase, as we will discuss the same effect is manifest when discussing quantities that are gauge invariant.
 } 
%
%


  Interestingly,  the equation of motion for the radial vector component is a quadratic algebraic equation, and 
   for some parameter
   choice  its roots can become complex. This  implies that  the vortex solution  ceases to exist in a  region
surrounding the origin of the radial coordinate, leading to a `thick'  singularity.  
   We will develop  a geometric understanding of this non-linear effect,
     and we will discuss some of its physical consequences.\footnote{
 Notice    that similar   obstructions to find complete solutions -- associated
   with the non-linearities of the equations --  are not novel to our system: other investigations of realization
   of Vainsthein mechanism in the context of massive gravity found similar behaviours \cite{Sbisa:2012zk}.}
   
   {
   It is also possible to study the backreaction  of the system when coupling it with gravity.
    The new gauge invariant field profiles excite new components of the energy momentum tensor,   that need
    to switch on appropriate metric components in order to satisfy Einstein equations.  On the other hand, we
    show that field dependent coordinate transformations exist, that `adapt' the geometry to our special
    vortex profile, and that make the resulting space-time manifestly axially symmetric.}

      


\section{A review: Vortex in the Abelian Higgs model}
 \subsection{The Abelian-Higgs system}
 
The Abelian Higgs Lagrangian is a U(1) gauge  invariant system  describing a massless gauge field,
$A_a$, and a self interacting complex scalar field $\Phi$. Within a mostly minus signature convention, 
it is given by 
\begin{equation}
\lag_{AH}[\Phi,A_a] =( D_a \Phi)^\dag D^a\Phi - \frac{1}{4}  F_{ab}  F^{ab} 
-\frac{\la}{4} \left (\Phi^\dag \Phi -\eta^2  \right)^2, \label{ahmlag}
\end{equation}
where $D_a =\partial_a +  i e A_a$ and $ F_{ab}= D_a A_b - D_b A_a$. Although our
  discussion can be  extended to four space-time dimensions,  we restrict ourselves to 2+1 dimensions labeled
  by $(t, \,x_1,\,x_2)$. Thus the mass dimensions of the quantities above are  $[\Phi\Phi^\dag]=[e^2]=[\lambda]=[\eta^2]=[A^2]=1$. We will come back later to
the Lagrangian in the form \eqref{ahmlag}, but in order to present vortex configurations it is
useful to take a shortcut by introducing the Ansatz
\bse\label{ansatz1}
\begin{align}\label{ansatz1a}
\Phi (x^\alpha)& = \eta X(x^\alpha) e^{i\chi(x^\alpha)}, \\
A_a(x^\alpha) &=\frac{1}{e} \left[ \hat A_a(x^\alpha) - \pd_a \chi(x^\alpha) \right]\label{ansatz1b},
\end{align}
\ese

\noindent where the new fields are all real and have mass dimensions $[\hat A_a]=1$ and $[X]=[\chi]=0$. 
The Lagrangian is invariant  under a U(1) gauge transformation 
\be
\Phi \to \Phi e^{i \xi} \hskip 1cm {\text{and}} \hskip 1cm 
 A_a \to A_a -e^{-1}  \partial_a \xi,\,\ee
  for an arbitrary function $ \xi$. The fields $X,$ and $\hat A_a$ introduced in
   eq. \eqref{ansatz1} are  invariant under such a  transformation.
   
   The advantage of using Ansatz \eqref{ansatz1} is that $\chi$ drops out
from the Lagrangian, that rewrites
\be
\lag_{AH}[X,\hat A_a] =  \eta^2 \pd_a X \pd^a X + \eta^2 X^2 \hat A_a \hat A^a - \frac{1}{4 e^2} F_{ab}F^{ab} 
- \frac{\la \eta^4}{4}\left( X^2 - 1\right)^2, \label{nogaugelag}
\te
where $F_{ab} = \pd_a \hat A_b - \pd_b \hat A_a$. In this way the gauge invariant,  physical degrees of freedom 
of the model have been extracted. 
 The corresponding
equations of motion
are 
\begin{subequations}
\begin{align}
\Box X - X \hat A_a \hat A^a + \frac{\la \eta^2}{2}\left(X^2 -1  \right)X &= 0, \\
\pd_a F^{ab} + 2 e^2 \eta^2 X^2 \hat A^b & =0.
\end{align}\label{noeq1}
\end{subequations}
We can identify two mass scales in these equations. The  scalar equation contains $\sqrt{2} m_X = \sqrt{\lambda}\eta$, while the last term in the vector equation contains a mass associated to
the vector field, $m_V = \sqrt{2} e \eta$. 
   As we will see next,  the mass scales $m_X$ and $m_V$ control
the field profiles, and therefore the localization properties  of the vortex.
\subsection{The Nielsen-Olesen vortex}\label{thevortex}
The vortex is a solution to gauge theories with scalar fields first studied by Abrikosov \cite{abrvortex}, and
 Nielsen and Olesen \cite{Nielsen:1973cs}. This  solution describes a vortex-like object
carrying a localized magnetic flux in its core. The size of the core depends on the mass scales of 
the theory, i.e. on the mass of the scalar and gauge fields. 

Some simplifying assumptions can be done. First,  $A_0$ is not a propagating 
degree of freedom, and it is consistent
 to set $A_0 = 0$ since this quantity appears at least quadratically in the Lagrangian. Furthermore, only static configurations are considered. Using Cartesian coordinates for the spatial part of the
 metric, the energy functional for such configurations is given by
\begin{equation} \label{intah1}
\mathcal E_{AH} = \int d^2x \left[ \frac{1}{2} B^2 + |\vec D\Phi|^2 + \frac{\lambda}{4}(|\Phi|^2-\eta^2)^2 \right],
\end{equation}
where $B=-F_{12}=\partial_{x_2}
 A_{x_1}- \partial_{x_1} A_{x_2} $, and  $\vec D$ has for components the spatial parts of $D_a$. Axial coordinates, that we  use more frequently, 
 are defined as usual as
 \bea
 r&=&\sqrt{x_1^2+x_2^2}
 \hskip1cm,\hskip1cm
 \theta\,=\,\arctan\left[\frac{x_2}{x_1}\right].
 \eea
  The 
integral \eqref{intah1}  spans over the entire space, therefore in order to keep the energy finite we require the
potential of the scalar field to vanish for large $r$,  thus $|\Phi|^2 = \eta^2$ asymptotically, or equivalently $X\to 1$ as $r\to\infty$.

\smallskip


\smallskip 
 Indeed
the potential minimum is isomorphic to a circle and has solutions characterised by
the phase $\chi$, $\Phi = \eta e^{i \chi}$. 
 Asymptotically, $\chi$ defines a mapping from a circle of radius $r$ in real space to 
a circle of radius $\eta$ in the complex plane. Mappings from one circle to $N-$circles 
are described by \be \label{apah}\chi = N \theta\,,\ee where $\theta$ is the polar
angle and the integer $N$ is the \emph{winding number}. It
counts the number of circles in complex space corresponding to a circle at spatial infinity. 
The winding number is a topological invariant, in the sense that  the asymptotic value of the phase \eqref{apah} cannot
be modified by a gauge transformation that is regular everywhere. Hence the winding number characterises different classes of finite energy solutions.

One further asymptotic condition that we must impose to keep the energy finite is $D_a \Phi =0$ at $r\to \infty$. 
 This requirement can be fulfilled thanks to the coupling of the scalar with the vector field, 
and it is crucial to avoid Derrick theorem.  Using the definition of $D_a$, this asymptotic
condition implies
\begin{equation}
A_a \sim \frac{i}{e}\partial_a \ln \Phi \sim -\frac{N}{e}\partial_a \theta, \label{boundary}
\end{equation}
as $r\to\infty$.
 Hence the vector field profile compensates for the scalar contribution at large $r$, keeping the energy finite \cite{Weinberg:2012pjx}. 
 Once  we know the asymptotic form of the gauge field for a vortex, we can readily compute its magnetic flux by integrating over a loop at $r\to\infty$,
\begin{equation}
{\text{Flux}}\, =\, e \int d^2 x B =-e \oint A_i dx^i = 2\pi N,
\end{equation}
then the flux is an integer multiple of $2\pi$. This result depends only on the asymptotic 
behaviour of the fields and therefore it holds also for the model of Galileon Higgs  that we use in the next section. 

 In addition to the asymptotic conditions at $r\to\infty$, we also need
\begin{equation}
A_a (r\to 0 )= 0, \ \ X(r\to 0)=0 \label{cond}
\end{equation}
 to guarantee the regularity of the solutions at the origin. 
As \eqref{nogaugelag} shows, when we use the Ansatz \eqref{ansatz1},  the Higgs
phase $\chi$ drops out 
  and 
the physical
degrees of freedom of the theory are only in $\hat A_a$ and $X$. In consequence, at this point we are free to fix
the  phase to depend on $\theta$ only, as 
 $\chi = N \theta$. After making this choice, we are no longer free to perform
a gauge transformation to remove the radial component of $ A_r$ (the longitudinal mode), 
though.  Nevertheless,  for the system we consider in this section, it is consistent to assume that  the component $ A_r$ vanishes.

More precisely, 
in order to {\it uniquely} determine the vortex  solution we impose the following two conditions \cite{Weinberg:2012pjx}:
\begin{enumerate}

\item Rotational symmetry, in the sense that the effects
of spatial rotations can be compensated by a  gauge transformation that is  uniform all over the space.

\item Invariance under reflection about the $x_1$ axis,
accompanied by complex conjugation of the scalar field. Such discrete symmetry  requires that the vector
components obey 
\be {A}_1(r,\,\theta)=-{A}_1 (r,-\theta) \hskip1cm {\text{ and}} \hskip1cm {A}_2(r,\,\theta)={A}_2 (r,-\theta).
\label{refco}\ee
\end{enumerate}

We introduce the following Ansatz for the gauge invariant  vector  components (recall that vector
 gauge invariant components are labeled with a hat)
\be
\hat{A}_i\,d x^i=\left[-\epsilon_{ij}\,\frac{{x_j}}{r} \,
\hat A_\theta(r)
+\frac{x_i}{r} \,\hat{A}_r(r)
\right]\,d x^i.
 \label{ansai}
\ee
The previous requirements implies that $X$,  $\hat A_\theta$, and $\hat{A}_r$   depend on $r$ only.
To make Ansatz \eqref{ansai} compatible with \eqref{refco} accompanied by a complex conjugation of the scalar field,
the second requirement  imposes that $\partial_r \chi\,=\,0$,  and at the same time $\hat{A}_r\,=\,{A}_r\,=\,0$.  This condition
is compatible with the equations of motion, since  $\hat{A}_r$  appears at least
quadratically in the action (this is a major difference with respect to the Galileon Higgs theory that 
we will discuss in the next section).

Within this Ansatz, the four equations given by \eqref{noeq1} are reduced to  two coupled equations  for the fields
$X$, $\hat A_\theta$:
\bse
\begin{align}
r \eta ^4 \lambda  X \left(1-X^2 \right)-\frac{2 \eta ^2 \hat A_\theta^2 X}{r}+2 \eta ^2 X'+2 r \eta ^2 X'' & =0,\label{novortexa}\\
-2 e^2 r \eta ^2 \hat A_\theta X^2 -\hat A_\theta'+r\hat A_\theta''&=0 .
\label{novortexb}
\end{align}\label{novortex}
\ese
Exact solutions to these equations can not be expressed in terms of standard functions; however several numerical and approximate results both for the Abelian Higgs model and  generalizations 
 exist   \cite[see e.g.][]{Lake:2010wt, Kawabe:1997ua, Jatkar:2000ei}. 

 \begin{figure}[!htb!]
\begin{center}
\includegraphics[width=0.45\textwidth]{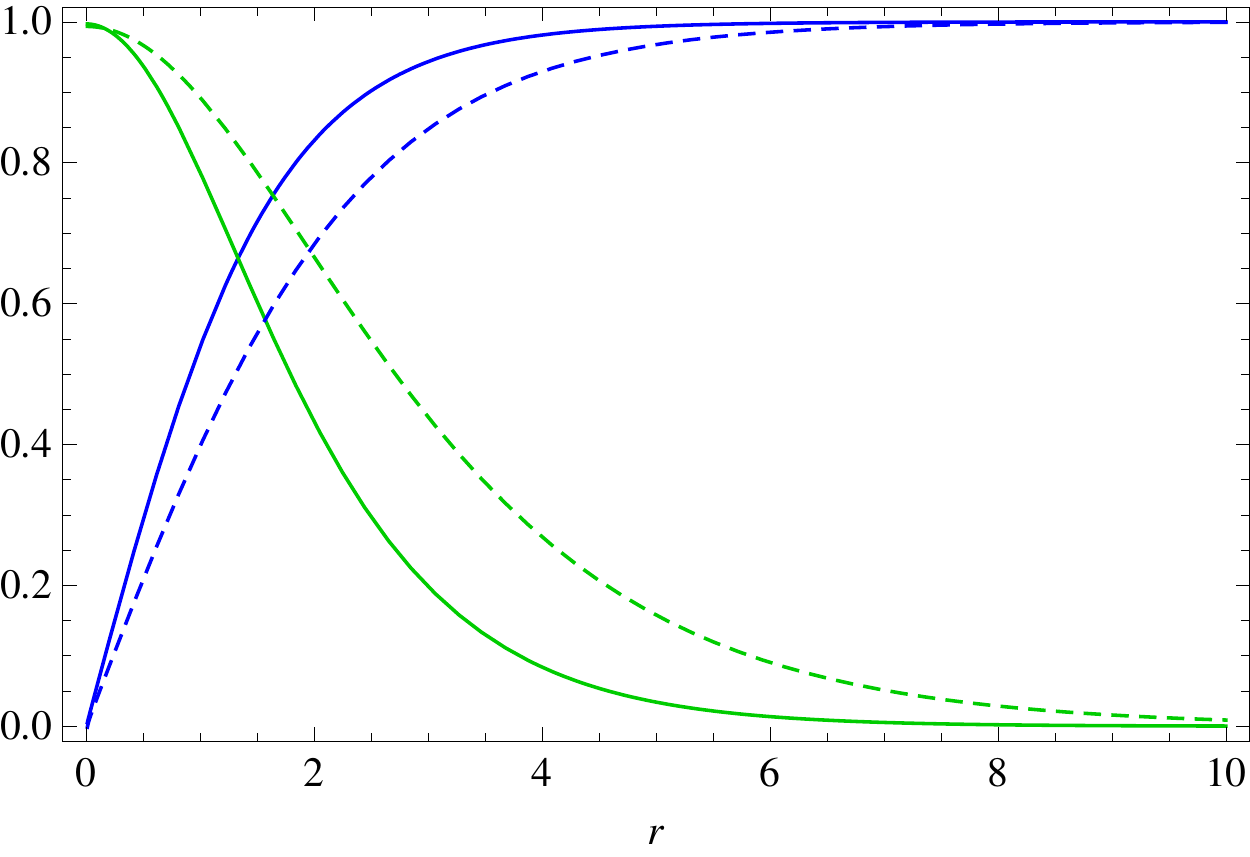}\caption{\it{$\hat A_\theta$ (green) and $X$ (blue) for Nielsen-Olesen vortices with $N=1$ and $\lambda=2 e^2$. For the solid lines $\lambda=\eta=1$, while for the dashed lines $\lambda=\eta=0.8$. By changing these parameters we are changing
the masses of the scalar and vector fields, and this modifies the width of the vortex core. }
}\label{novorfig}
\end{center}
\end{figure}

 In Fig. \ref{novorfig} we show the representative example of a numerical solution to the
equations \eqref{novortex}, it corresponds to a vortex with winding number $N=1$. Notice
that although the equations of motion do not depend explicitly on the vorticity, 
the boundary conditions do so (see eq. \eqref{boundary}). In this sense,  the winding number  does control the  solutions.
 The profile for the scalar and magnetic fields can be understood as follows \cite{Weinberg:2012pjx}. Since there is non-vanishing
 vorticity, there must be regions where the magnetic field is different from zero.  The gauge field acquires a mass
 when the scalar field is non-vanishing. Hence it is energetically favoured for the magnetic field to be concentrated in a region near the origin, 
  where the scalar field acquires a value close to zero; moreover, rendering this region thicker   reduces the  magnetic energy contribution.   
 Contrasting  this effect, and favouring  a smaller vortex core, is the fact that it costs energy for the scalar to be away from the minimum of its  Mexican hat 
 potential. The relative strength of these two competing effects is determined by $\lambda/e^2$.

\bigskip 
\subsection{Boundedness of the Hamiltonian and the BPS bound}\label{boundah}
Another 
 important property of NO configurations 
  is the existence of a particular point in
the parameter space known as the \emph{Bogomol'nyi point} \cite{Bogomolny:1975de}, or alternatively as the BPS bound, given by $\lambda=2 e^2$. To see what 
makes this point special we need to consider the  energy functional for static configurations:
\begin{equation}
\mathcal E_{AH} = \int d^2x \left[ \frac{1}{2} B^2 + |\vec D\Phi|^2 + \frac{\lambda}{4}(|\Phi|^2-\eta^2)^2 \right].
\end{equation}
 Noticing
that $|\vec D_\Phi|^2 = |(D_1 \pm i D_2)\Phi|^2 \mp e B |\Phi|^2 \pm 2 \pd_{\{i} J_{j\}}$, where
$2 i J_j = \Phi^\dag D_j \Phi - \Phi (D_j\Phi)^\dag$, making a `complete the square' argument and 
dropping boundary terms, the energy functional becomes
\begin{equation}
\mathcal E_{AH} = \int d^2 x\left[\frac{1}{2}\left( B\mp e (|\Phi|^2 - \eta^2) \right)^2 + |D_{\pm}\Phi|^2 + \left(\frac{\lambda}{4} - \frac{e^2}{2} \right) \left( |\Phi|^2-\eta^2 \right)^2 \mp e \eta^2 B  \right].
\end{equation}
This expression shows that it is bounded from below. 
The potential term  vanishes if $\lambda = 2e^2$. The other conditions to minimize the energy functional  are
\begin{subequations}
\begin{align}
D_{\pm}\Phi &= 0, \\
B & = \pm e \left(|\Phi|^2 - \eta^2 \right).\label{sdeqb}
\end{align} \label{sdeq}
\end{subequations}
These are known as the BPS self-duality equations \cite{Prasad:1975kr, Coleman:1976uk}. 
By considering axially symmetric configurations these equations reduce to \eqref{novortex} with $\lambda = 2 e^2$. 

\section{The vortex in presence of  Galileon Higgs interactions}

 \subsection{The Higgs model including higher order derivative self-interactions}

The Abelian Higgs model defined by Lagrangian \eqref{ahmlag} can be extended 
including 
 non-linear derivative self-interactions of the
gauge field, 
  which can be relevant in the
context of   vector-field models for dark energy 
 (see, e.g. \cite{Clifton:2011jh} for a general review). 
 Such derivative interactions are ghost free and gauge invariant; since they
 involve covariant derivatives they couple the Higgs to gauge fields. They  have been introduced in 
  \cite{Hull:2014bga}
  as a way for Higgsing the Abelian symmetry breaking model of vector Galileons \cite{Tasinato:2014eka, Heisenberg:2014rta,Tasinato:2014mia}.
   Regardless of this motivation, they can be seen as ghost-free derivative extensions of the Abelian Higgs model, 
   related with 
 Galileon  systems in an appropriate
 decoupling limit (as we will review below). 
  This connection with Galileons can be useful for analysing the (non-)renormalization properties of our derivative interactions under quantum
  corrections. This is an issue outside the scope of this work, and that we leave 
 for further studies. 
  
   In three space-time dimensions
only  two  of the three new proposed operators can be defined,
  and in static situations only the lowest dimensional of these operators is different from zero. This operator, which we call $\mathcal L_6$ in reference to its mass dimension, is given by
\bea
  \label{defol}
	\mc{L}_{\mathrm{6}} &= &
	-\frac{1}{\Lambda^3}\,\ve^{
	c
	a_{1}a_{2}}\ve_{cb_{1}b_{2}}\,\left[\, \al_{(6)} L_{a_{1}}^{\,b_{1}}P_{a_{2}}^{\,b_{2}}+\beta_{(6)}L_{a_{1}}^{\,b_{1}}Q_{a_{1}}^{\,b_{2}}\,\right],
%
\eea
where $\Lambda$ has dimensions of mass and $\al_{(6)}$ and $\beta_{(6)}$ are dimensionless; all these parameters  are constant. $\ve_{abc}$ is a totally antisymmetric tensor in three dimensions with  $\ve_{123}=1$,
 and the gauge invariant operators $P_{ab}$ and $Q_{ab}$ are expressed in terms of the Higgs covariant derivatives  by
\bse
 \begin{align}
	L_{{a}{b}}&\equiv \frac12 \left
	[({D}_{{a}}\Phi)^{*}({D}_{{b}}\Phi)+({D}_{{b}}\Phi)^{*}({D}_{{a}}\Phi)\right]\,,
	 \\ P_{{a}{b}}&\equiv  \frac12 \left[\Phi^*{D}_{{a}}{D}_{{b}}\Phi +\Phi\,\left(
	 {D}_{{a}}{D}_{{b}}\Phi\right)^* \right]\,,\\
	  \,Q_{{a}{b}} &\equiv \frac{i}{2} \left[
	  \Phi \left( {D}_{{a}}{D}_{{b}}\Phi\right)^* 
	  -\Phi^*{D}_{{a}}{D}_{{b}}\Phi 
	  	  \right]\,. 
\end{align}
\ese
These operators are symmetric as can be verified by expanding the gauge derivatives.  Following
the same route as in the previous section, we proceed to write down the Lagrangian in terms
of the fields $X, \chi$ and $\hat A_a$. Using the ansatz \eqref{ansatz1} we write the previous operators in terms of the gauge invariant field $\hat A_a$ and the real scalar field $X$,

\bse
\begin{align}
	L_{{a}{b}}&= \eta^2\partial_{{a}}X \partial_{{b}}X + \eta^2 X^2 \hat A_{{a}}\hat A_{{b}}\,,\\
	P_{{a}{b}}&= \eta^2 X \partial_{{a}}\partial_{{b}}X -\eta^2  X^2\hat A_{{a}}\hat A_{{b}} \,, \\Q_{{a}{b}} \, 
	  &= \frac{1}{2}\eta^2 \,[\partial_{{a}}(X^2\hat A_{{b}})+\partial_{{b}}(X^2\hat A_{{a}})]\,.
\end{align}
\ese
Notice that the phase $\chi$ does not appear in the previous expressions, as
expected since all quantities are written in a gauge invariant form.   
For simplicity, we  focus only in the part of the Lagrangian proportional to $\beta_{(6)}$ that 
 depends on vector field derivatives, and that as we will discuss switches on 
  new field profiles
    that we wish to investigate.  Hence, making a rescaling and redefining 
$\beta_{(6)}=\Lambda^3 \beta$ the Lagrangian $\mc{L}_{6}$ that we study  is
\begin{equation}
\mc{L}_{6}[X,\hat A] =  \beta \eta^4\left(\partial_a X\partial_b X + X^2 \hat A_a \hat A_b\right)\left[\eta^{ab}\partial^{c}(X^2 \hat A_c) - \partial^a (X^2 \hat A^b) \right].\label{lag6}
\end{equation}

\subsection{(Bi)galileons from decoupling limit}

In this subsection, we review the connection between the Higgs derivative self-interactions that we consider,
and Galileon theories, referring  to \cite{Hull:2014bga} for a more extensive discussion. 
 
 We 
  first
need to  expand the Higgs around its {\it vev}: using our notation, this implies that we  introduce  the field $h$ as  a perturbation around the $X=1$ Higgs {\it vev}: 
\be
X\,\equiv\,1+\frac{h}{\sqrt{2}}.
\ee

The total Lagrangian, expanded in terms of the field $h$ and the vector fields, reads

\bea \label{tlas1}
{\cal L}_{tot}&=&-\frac{1}{4} F_{\mu \nu} F^{\mu \nu}- m_A^2\, \hat{A}^2-\tilde \beta\,\hat A_\mu \hat A^\mu\,\partial_\rho \hat A^\rho
\nonumber
\\
&&- \frac12\,(\p h)^2 -
\frac12\,m_h^2\,
 h^2\,-\frac{\sqrt{\lambda}\,m_h}{4}\,h^3\,-\frac{{\lambda}\,}{{16}}\,h^4
 -\sqrt{2}\,e\,m_A\,h\,\hat{A}_{\mu} \hat{A}^{\mu}-\frac{e^2}{2}\,h^2\,\hat{A}_{\mu} \hat{A}^{\mu}
 \,
\nonumber
\\
&&+\frac{4\,e\,\tilde \beta}{3\,m_A}\,\left(\sqrt{2}\,h +\frac{3\,e}{2\,m_A}\,h^2\, +\frac{e^2}{\sqrt{2}\,m_A^2}\, h^3+\frac{e^3}{8\,m_A^3}\,h^4\right)\,\left(\hat A_\mu\,\hat A^\nu\,\partial_\nu \hat A^\mu-\hat A_\mu\,\hat A^\mu\,\partial_\rho \hat A^\rho \right)
\nonumber
\\
&&+\frac{\tilde \beta }{3\,m_A^2 }\,\left(  1+\frac{\sqrt{2}\,e}{m_A}\,h+\frac{e^2}{2\,m_A^2}\,h^2\right)\left(\partial_\mu  h \,\partial^\nu  h \,\partial_\nu \hat A^\mu-
\partial_\mu  h\, \partial^\mu  h \,\partial_\rho \hat  A^\rho
 \right)\,, \label{explag}
\eea
with
\bse
\bea
m_A&=& e \,\eta\,,
\\
\tilde{\beta}&=&-\frac{3 \,e^3\,\beta_{(6)} \,v^4}{2 \Lambda^4} \label{deftb}
\,,
\\
m_h&=&\sqrt{\lambda}\,\eta\,.
\eea
\ese

Such Lagrangian is gauge invariant; nevertheless the system 
exhibits spontaneous symmetry breaking, and 
it  contains a mass for the Higgs field $h$ and the
vector field, as well as various derivative interactions
 between the Higgs and the vector components. Since the vector is now massive, it propagates three degrees
 of freedom, two transverse and one longitudinal. 
 We are now interested to exhibit a `decoupling limit' where the only interactions left are the ones between
 the Higgs with itself and with the longitudinal component of the vector. We will learn that such interactions have a bi-Galileon structure. 
 
 In order to make such interactions more manifest, we introduce by hand  a `St\"uckelberg' field
 to identify more simply the vector longitudinal mode:    whenever
 we meet a vector in the Lagrangian \eqref{tlas1} we substitute it  with 
  \be
  \hat{A}_\mu\,\to \, \hat A_\mu- \frac{\partial_\mu\,\hat{\pi}}{\sqrt{2}\,m_A}.
  \ee
  The theory acquires an additional gauge symmetry $A_\mu\to A_\mu+2 m_A\,\omega$, $\pi\to\pi+\omega$.  Choosing a gauge in  which $\pi=0$ one recovers the original Lagrangian.  The field $\pi$ plays the physical role of the vector
  longitudinal polarization.
  
  Consider the decoupling limit 
 \be \label{limit1}
e\to 0\hskip0.5cm,\hskip0.5cm \lambda\to 0\hskip0.5cm,\hskip0.5cm \beta_{(6)}\to 0\hskip0.5cm,\hskip0.5cm \eta\to \infty\,\,,
\ee
such that 
\be \label{limit2}
m_A\to0\hskip0.5cm,\hskip0.5cm m_h\to0\hskip0.5cm,\hskip0.5cm \tilde \beta \to 0\hskip0.5cm,\hskip0.5cm
 \frac{\tilde \beta}{m_A^3}\,=\,{\rm fixed}\,\equiv \,\frac{1}{\Lambda_{g}^3}\,\,,
\ee
where $\Lambda_{g}$ is a mass scale  associated with   the Galileon interactions.
 In order to have a correctly normalized kinetic term for the St\"uckelberg  field $\pi$ (corresponding
 to the vector longitudinal polarization) we   rescale it,  and define $ \pi=\hat \pi/(\sqrt{2} m_A)$.
  Within the decoupling limit, $\hat \pi$
   plays the role of Goldstone boson of the broken symmetry. 
   In the limit (\ref{limit1},  \ref{limit2}), when expressed in terms
  of the canonically normalized Goldstone field $\hat \pi$, the total Lagrangian ${\cal L}_{tot}$ reduces to 
  \be
  {\cal L}_{tot}\,=\,-\frac14 F_{\mu \nu} F^{\mu \nu}
  -\frac{1}{\Lambda_g^3} \,\left(\partial_\mu \hat \pi \partial^\mu \hat \pi\right) \,\Box \hat \pi-\frac{1}{3\,\Lambda_g^3}
  \left( \partial_\mu  h\, \partial^\mu  h \,\Box \hat \pi-
  \partial_\mu  h \,\partial^\nu  h \,\partial_\nu \partial^\mu \hat \pi
  \right)\,.
  \ee
Hence  in this decoupling  limit the Lagrangian acquires a bi-Galileon structure, since the  Higgs 
 itself acquires bi-Galileon couplings with the field $\hat \pi$, corresponding
 to the vector longitudinal polarization. Outside the decoupling limit, the Higgs couple with the transverse
 polarizations of the vector as well, and this fact allows the system to circumvent Derrick's theorem and to  find vortex solutions of finite energy.

\subsection{Equations of motion for a vortex configuration}

As for the Abelian-Higgs model, the phase $\chi$ -- and the vorticity as well  --
 do not appear explicitly in the equations of motion, that
 can be expressed in a gauge invariant form.
At large $r$ we  impose the phase to be asymptotically
 \be \chi = N\theta,\ee
   so to equip the configuration  with a topologically invariant
 winding number.

On the other hand,  analogously
to the case of the NO vortex, the  degrees of freedom $\hat A_a$ and $X$ are aware of the value of the vorticity  through the boundary conditions  \eqref{boundary}, which together with $|\Phi|^2= |\eta^2|$
guarantee that the static energy functional is finite since all the terms in the Lagrangian vanish
asymptotically. Hence the same asymptotic conditions for
a Nielsen-Olesen vortex in the Abelian Higgs model remain valid when $\mathcal L_6$ is turned on. The presence
of the vector 
allows us to find finite energy vortex configurations.\footnote{ 
 This does not necessarily  mean that ours are the most general conditions to get finite energy solutions, since
non-trivial cancellations might occur between the derivative terms in $\mathcal L_{AH}$ and those in $\mathcal L_6$.  However we do not consider this possibility in this work.} In terms of our Ansatz of eqs. \eqref{ansatz1} and \eqref{ansai}, that
we write again here 
\bse\label{ansatz3}
\bea
\Phi & =& \eta X(r) e^{i\chi}\,,
\\
A_a(x^\alpha) &=&\frac{1}{e} \left[ \hat A_a(x^\alpha) - \pd_a \chi(x^\alpha) \right] ,
\\
\hat{A}_i\,d x^i&=&\left[-\epsilon_{ij}\,\frac{{x_j}}{r} \,
\hat A_\theta(r)
+\frac{x_i}{r} \,\hat{A}_r(r)
\right]\,d x^i,
 \label{ansai2}
\eea
\ese
the 
asymptotic conditions required in order to get finite energy solutions are
\begin{equation}
\hat A_\theta \to 0, \ \ \hat A_r \to 0, \ \  X\to  1,
\end{equation}
at  $r \to \infty$. { Additionally, in order to compensate for the scalar contributions to the energy density, recall that $A_\theta$ satisfies the asymptotic condition \eqref{boundary}}.

The equations of motion can be expressed in fully covariant form. 
 However it is more convenient for our purposes  to write $\mathcal L_6$ in terms of the components of  Ansatz  \eqref{ansai2}
\begin{align} \label{lag6comp}
\mathcal L_6 = & \frac{4\beta\eta ^4}{r^3} X{}^2 \left[r^2 \hat A_r{} {}^3 X{}^2+r \hat A_\theta{}{}^2 X{}^2 \hat A_r{} '{} + \hat A_r{} {} \left(2 \hat A_\theta{}{}^2 X{}^2-r \hat A_\theta{}{} X{}^2 \hat A_\theta{}'{}+r^2 X'{}^2\right)\right],
\end{align}
 and to
compute the equations of motion explicitly by taking the variation of the action with respect to these components. In doing so
we should be careful  not to oversimplify things. In particular, since none of our fields depends on time we can, and we had, set the time 
component $\hat A_0$ equal to zero, but we cannot do the same for the radial component, $\hat A_r$, because \eqref{lag6comp} contains terms linear in $\hat A_r$ whose contribution to the equations of motion would be missing if we set $\hat A_r=0$.  

As a consequence,  the complete Lagrangian
\begin{equation}
\mathcal L_{AHG} = \mathcal L_{AH} +\mathcal L_6, \label{lahg}
\end{equation}
leads to  {\it three}  independent equations of motion, expressed in terms of gauge invariant quantities:  
\bse\label{sys8}
\begin{align}
r \eta ^4 \lambda  X \left(1-X^2 \right)-\frac{2 \eta ^2 \hat A_\theta^2 X}{r}+2 \eta ^2 X'+2 r \eta ^2 X''  & \nn \\ 
 + 16 \beta \eta^4 X^3 \hat A_r  \left[\hat A_{r}^2 +\frac{2{}  \hat A_{\theta}^2}{r^2}+\frac{ {} \hat A_{\theta}^2  \hat A_{r} '{}}{r \hat A_r}-\frac{ {}  {} \hat A_{\theta}{}  \hat A_{\theta}'{}}{r}-\frac{ {} \hat A_{r} '{} X'{}}{2 X \hat A_r}-\frac{ {} {} X'{}^2}{2 X^2}-\frac{ {} {} X''{}}{2 X}\right]&=0,\label{sys8a} \\
 -2 e^2 r \eta ^2 \hat A_\theta X^2 - \hat A_\theta'+r\hat A_\theta'' 
+12 \beta e^2 \eta^4 X^4  r  \hat A_\theta  \left[\frac{ {} \hat A_{r} {} }{r}+{} \hat A_{r} '{}+\frac{4 {} \hat A_{r} {}  X'{}}{3 X}\right]&=0,\label{sys8b}\\
- r \eta ^2 \hat A_r {} X{}^2+6 \beta \eta^4 X^4 \left[\hat A_{r}^2+\frac{ \hat A_{\theta}^2}{r^2}-\frac{ \hat A_{\theta}{} \hat A_{\theta}'{}}{r}-\frac{4 {} \hat A_\theta^2  X'{}}{3 X r}+\frac{ {} X'{}^2}{3 X^2}\right]&=0.\label{sys8c}
\end{align}
\ese 
We reiterate that, as for the NO vortex, the phase and the vorticity do not appear in the equations of
motion for our gauge invariant quantities;  but they do determine the profile  of the fields by means of the boundary conditions
required to get finite energy solutions.  
  
\begin{figure}[!htb!]
\begin{center}
\includegraphics[width=0.45\textwidth]{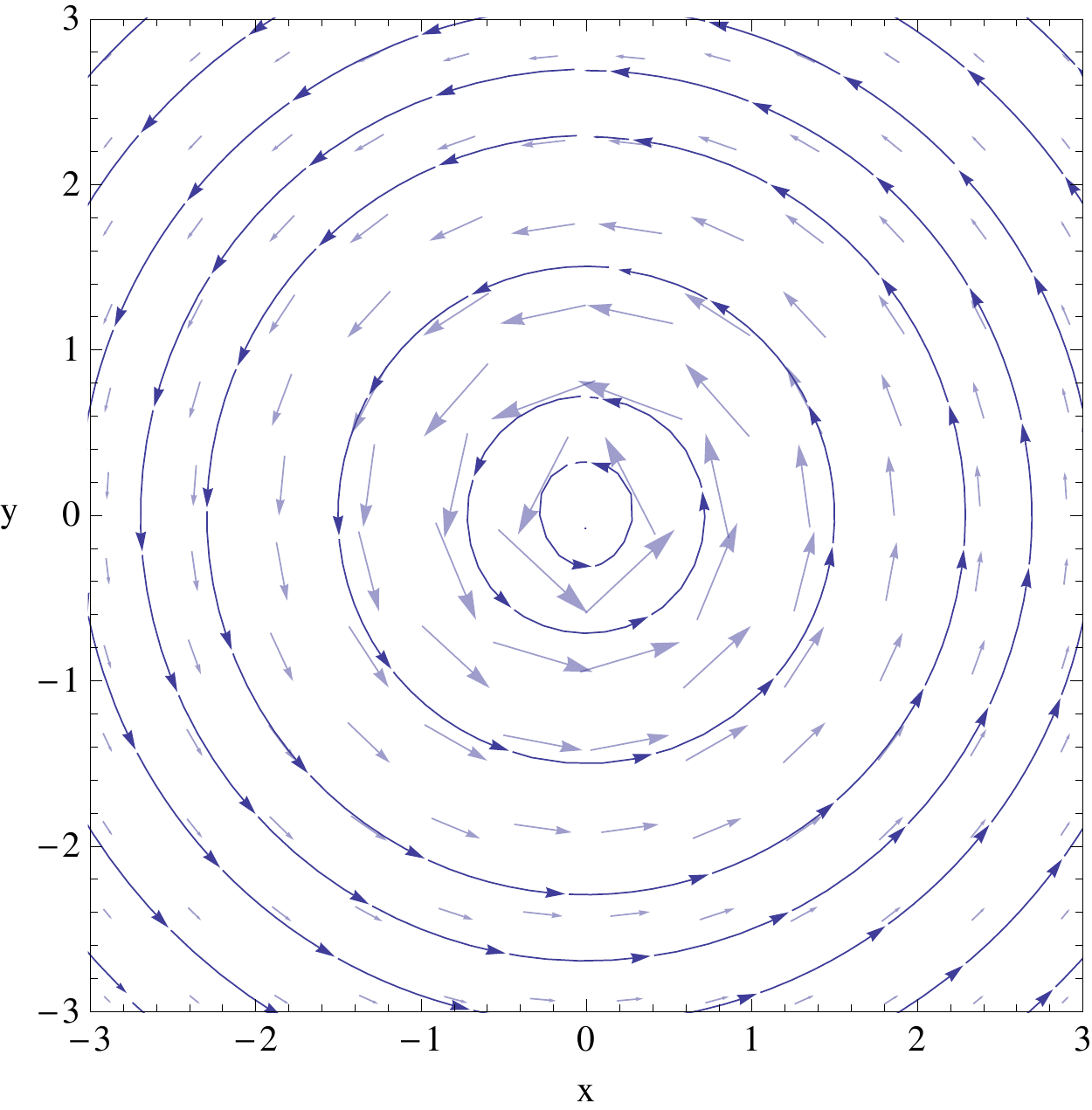} \ \includegraphics[width=0.45\textwidth]{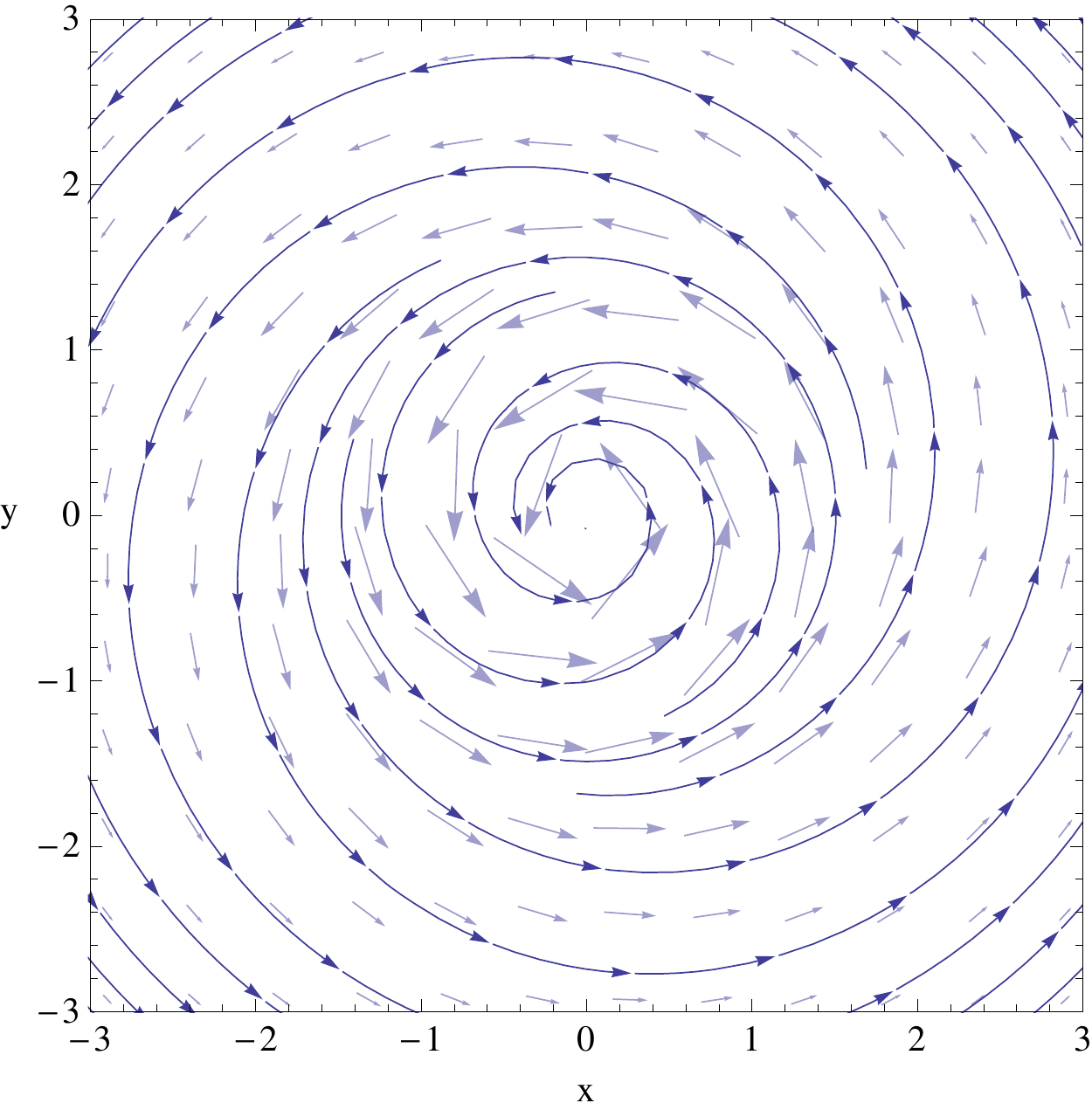} \caption{{\it A representative example of the gauge invariant vector field $\hat A_i$ in Cartesian coordinates for a NO vortex (left) and  for a 
Galileon Higgs vortex with $\beta=0.40$. Reflection invariance around a Cartesian axis  is lost in the Galileon Higgs case.}}\label{break1}
\end{center}
\end{figure}

The  most interesting new
 feature of this set of equations is eq. \eqref{sys8c}, the algebraic equation of motion for $\hat A_r$. For $\beta\neq 0$ this quadratic
  algebraic  equation does not admit the solution $\hat A_r=0$. Instead the  formal solution of this equation --  compatible with our asymptotic
  conditions for vanishing asymptotic gauge fields  -- is
  \begin{equation}
\hat A_r = \frac{r}{12 \beta\eta^2 X^2 } \left[1-\sqrt{1-\left(\frac{12 \beta \eta^2 X^2}{r}\right)^{2}\,\left[\frac{\hat A_\theta}{r^2}( \hat A_\theta -  r    \hat A_\theta'{})+\frac{X'}{3X }\left( \frac{ X'{}}{X} - \frac{4 \hat A_\theta^2  }{r} \right)\right]}\right]. \label{rootm}
\end{equation}

A non-vanishing $\hat A_r$ implies that we are violating the second requirement discussed in the previous section (in
particular eq. \eqref{refco}), hence we have a system that is not invariant under reflection around the $x_1$-axis, accompanied by a complex conjugation of the scalar field
 \footnote{ Notice that, by switching on a radial vector component,  we are focussing on a particular pattern of breaking the rotational and discrete 
  symmetries  of the NO system. Other possibilities might exist -- for example by considering an Ansatz with explicit dependence on the angular coordinates -- but
 we will not consider them in this work.
 }. 
The expression \eqref{rootm} for $\hat A_r$ can be substituted into eqs. 
  \eqref{novortexa} and \eqref{novortexb} to find solutions for $X$ and $\hat A_\theta$. The solutions for $\hat A_\theta$,
  $\hat A_r$
can be included into the Ansatz   \eqref{ansai} to obtain  configurations for the gauge invariant  components $\hat A_1$ and $\hat A_2$ in cartesian coordinates. Fig \ref{break1}
  shows a comparison of the profiles for $\hat A_1$ and $\hat A_2$ between NO and
our vortex solution. The breaking of reflection symmetry is evident. { We emphasize that in Fig    \ref{break1}  we are plotting gauge invariant, physical quantities. Our resulting vortex configurations do not switch
  on new electric fields, but nevertheless they locally modify the profiles for the Higgs and magnetic fields associated with NO solutions. 
}
    
      Notice that the configuration
  \eqref{rootm} contains
  a square root --  being solution of a quadratic equation --  hence for some choices of parameters a real solution
  for $\hat A_r$
   might not exist in some regions of the 
   radial coordinate. This fact is crucial for determining explicit solutions: we discuss this issue in what comes next.

\subsection{Constructing Galileon Higgs vortex solutions }\label{exsol}

\begin{figure}[!htb!]
\begin{center}
\includegraphics[width=0.55\textwidth]{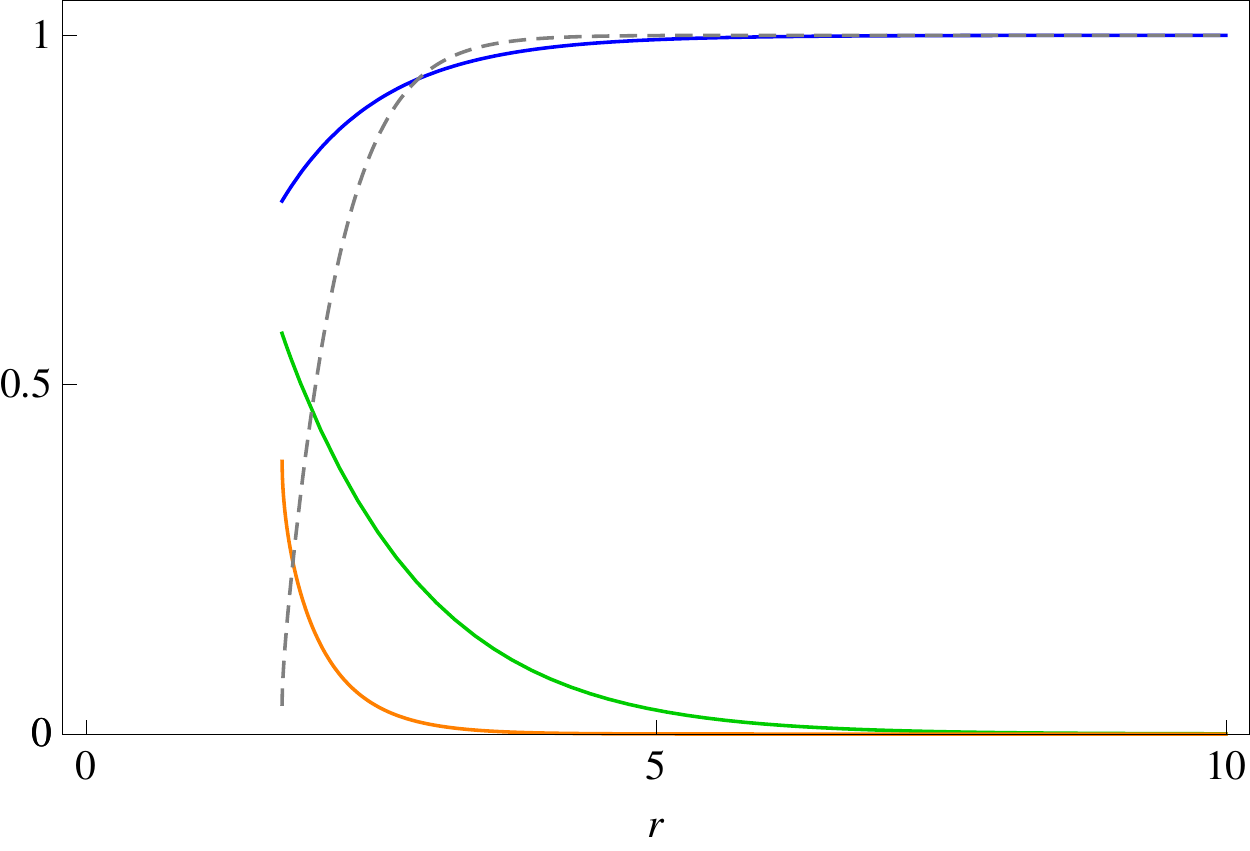}\caption{\it{$\hat A_\theta$ (green), $\hat A_r$ (orange)
and $X$ (blue)  for a solution to eqs. \eqref{sys8} with $\beta=0.5$, $\lambda=1$ and $\eta=1$. This
 solution breaks down at the point where the argument of the square root in \eqref{rootm}
 (dashed line) drops to zero, showing that $\hat A_r$ does not admit a real solution for every
set of parameters.}
}\label{break}
\end{center}
\end{figure}

From the equations of motion for the AHG (Abelian Higgs Galileon) system, \eqref{sys8},  we learn that although $\hat A_r$ does not have a dynamical equation of motion, it constrains the space of solutions to a subset for which $\hat A_r$ is real.

 To see that this
is indeed a constraint, we start 
 considering an extreme, singular example in  Fig. \ref{break}, where we present an explicit solution for our equations with $\beta=0.5$,
 selecting
  all the
parameters and boundary
  conditions equal to those of a NO vortex.
 Although this solution is well behaved asymptotically, it breaks down before the vortex is formed. This occurs when $\hat A_r$ becomes complex, and is associated
 with the formation of a singularity. {  This example indicates clearly that in order to avoid singularities and find regular vortex solutions, we will have to control
   some features of our configuration.}

\begin{figure}[!htb!]
\begin{center}
\includegraphics[width=0.45\textwidth]{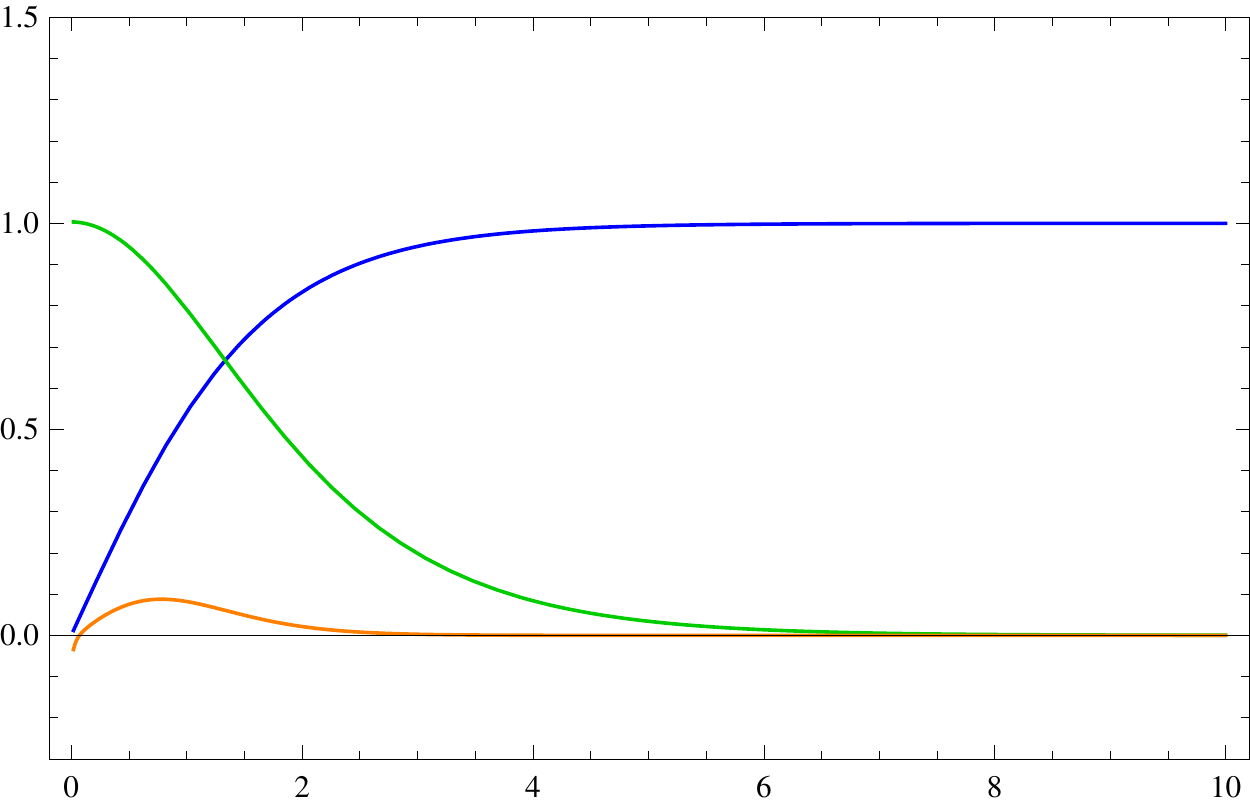} \ \includegraphics[width=0.45\textwidth]{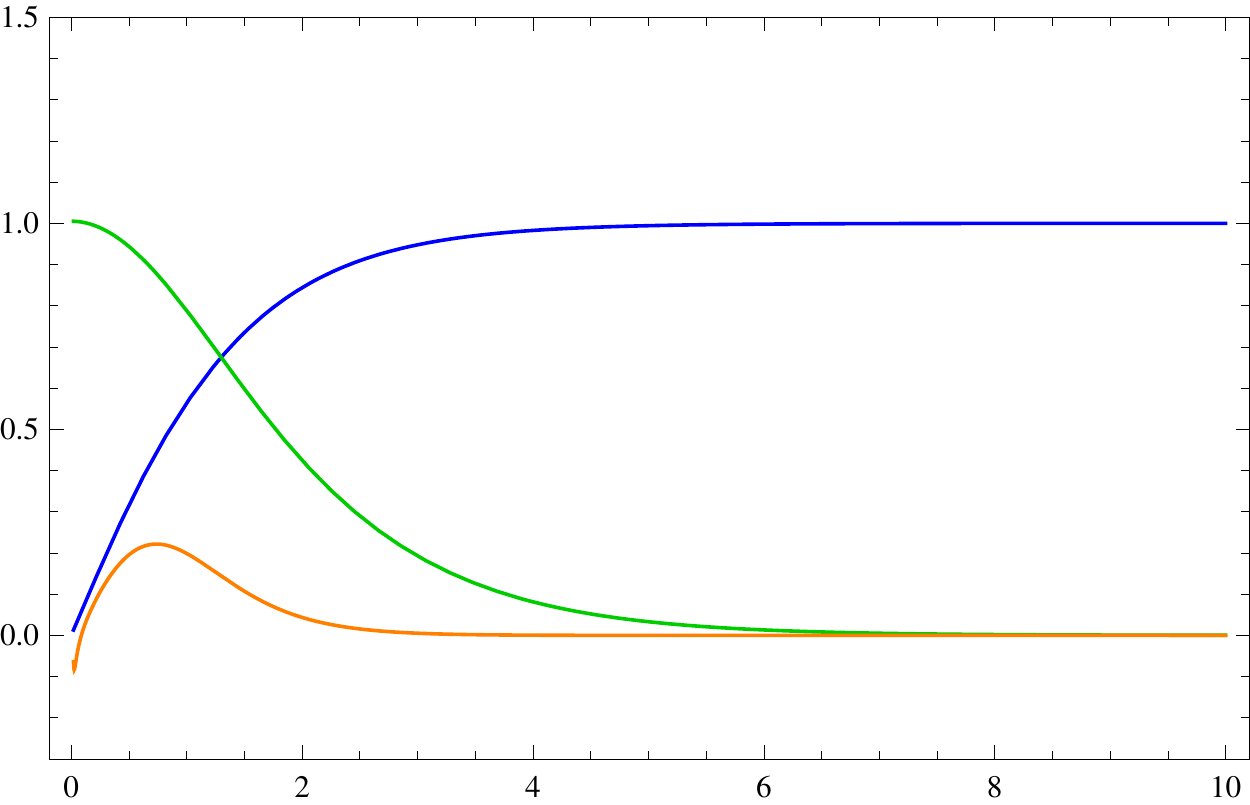}  \\
\includegraphics[width=0.45\textwidth]{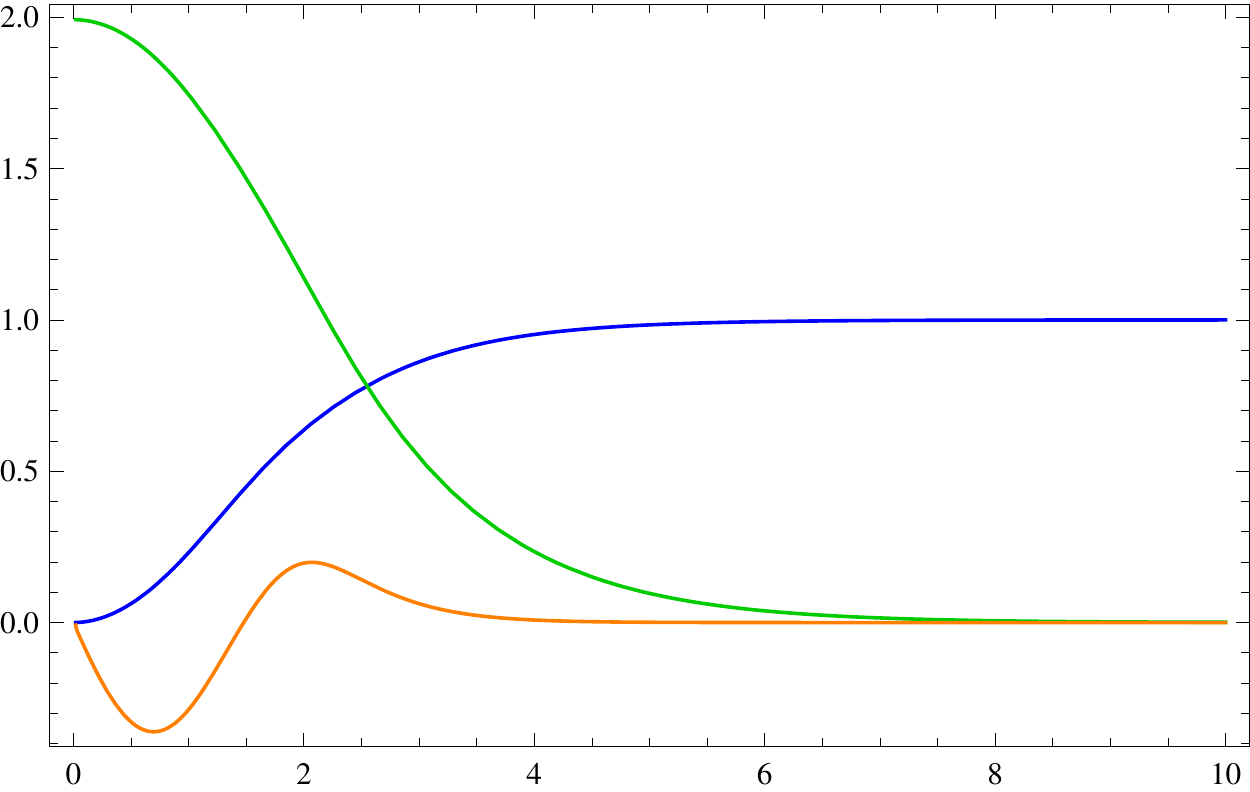} \ \includegraphics[width=0.45\textwidth]{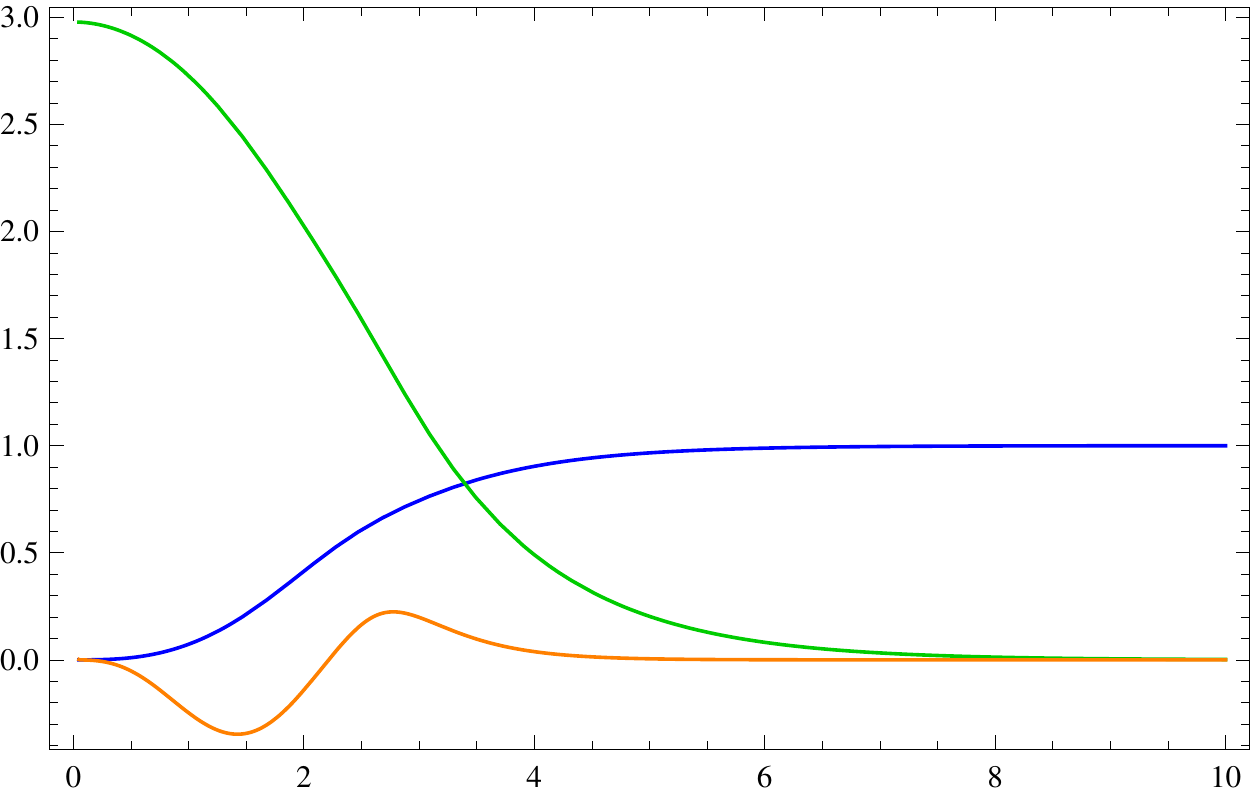} \caption{\it{From top left to bottom right: $\beta = 0.10\,\, (N=1), \,\beta = 0.20\,\, (N=1) , \,\beta = 0.40\,\, (N=2) \text{ and }  \,\beta = 0.49 \,\,(N=3)$. The fields shown are  $\hat A_\theta$ (green), $\hat A_r$ (orange)
and $X$ (blue). In all cases $\lambda=\eta=1$.} 
}\label{sols}
\end{center}
\end{figure}

To incorporate numerically the restriction imposed by $\hat A_r$, we allow the boundary conditions to vary until we find regular solutions across all the space. The system of equations
that we solve is obtained by substituting  \eqref{rootm} into \eqref{sys8a} and \eqref{sys8b}.  In Fig. \ref{sols} we show some of the solutions for different values of the coupling
constant $\beta$. The
boundary conditions for `large $r$' are imposed at $r=20$. Note that $\hat A_{\theta}(r\to\infty)$ does not
contain direct information about the vorticity since asymptotically such information cancels
out in the gauge transformation \eqref{ansatz1b}.  However, at a finite but large $r$ the
cancellation is not exact, and $\hat A_\theta$ is affected by the value of the vorticity.  Conversely,  a change in the boundary conditions for $\hat A_\theta$ and its derivative implies a change in the vorticity. In view of \eqref{ansatz1b}
and \eqref{cond}, the change of vorticity becomes manifest at $r= 0$, since $\hat A_\theta(r\to 0) = N$.
This is indeed seen in  Fig. \ref{sols}, for example the vortices with $\beta=0.4$ and $\beta =0.5$   correspond 
approximately to $N=2$ and $N=3$ respectively. Increasing the vorticity, we   find well-behaved solutions along  
the entire radial direction, even for larger values of the parameter $\beta$.

\begin{figure}[!htb!]
\begin{center}
\includegraphics[width=0.65\textwidth]{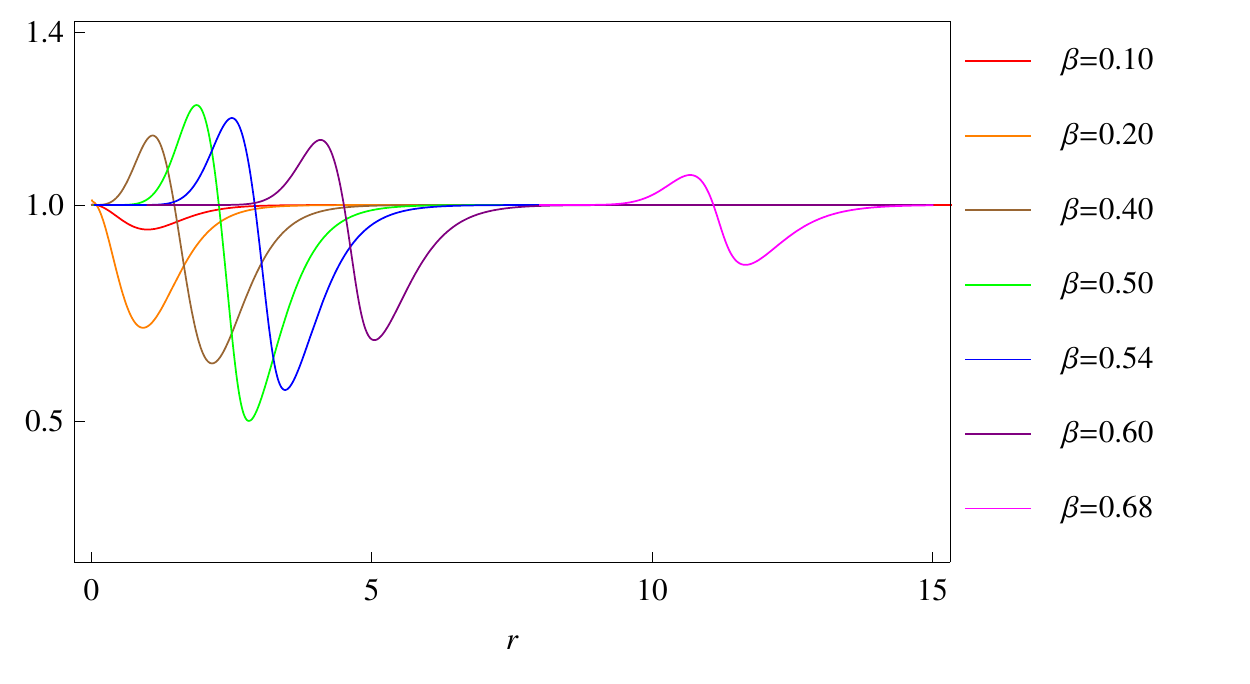}  \caption{\it{Argument of the square root that appears in $\hat A_r$, equation \eqref{rootm}. The deviations from one correspond to regions
where the non-linear effects can become  large. 
}}\label{arg} 
\end{center}
\end{figure}

 %
%
%
After analysing several numerical solutions with different boundary conditions, we conclude
that there is a minimal vorticity to obtain complete solutions in  the entire radial direction. 
 Such minimal value   increases with $\beta$ in a non-linear way. For example, for $0 <\beta \lesssim 0.25$ any vorticity is allowed, for $0.25 \lesssim \beta \lesssim 0.44$ the minimal vorticity
is $N=2$ and for $0.44 \lesssim \beta \lesssim 0.50$ the minimal vorticity is $N=3$.

 
 The fact that  increasing vorticity one finds 
 real solutions over all the radial coordinate might be interpreted
 as follows. As explained at the end of Section \ref{thevortex}, a vortex configuration is
 a balance between  the tendencies of the magnetic field  to get localized near the origin, and of the scalar
 field to lie on the minimum of its Mexican hat potential.  Increasing vorticity changes
 the boundary conditions for the gauge field,  and  causes the vortex to become wider.
  When the Higgs derivative self-interactions are turned on, they can destabilise  the aforementioned balance, since the new 
  contributions of the gauge component $\hat A_r$ tend to make the the field profiles wider, up to a point where no static configurations exist.
  A way out 
 is to increase the vorticity, 
  since changing boundary conditions for the gauge field do allow for a wider vortex configurations, that are able to accommodate sizeable contributions of
    $\hat A_r$.



 The argument of the square root in the algebraic solution for $\hat A_r$, \eqref{rootm}, is shown in Fig. \ref{arg}.  If such profiles are close to the value $1$, then
 $\hat A_r$ is close to zero.  
 The profiles support
the interpretation given above for the existence of a minimum vorticity.  
 We see that the $\hat A_r$ profile acquires a sizeable `bump'   at a distance from the origin that increases with $\beta$.  Such non-trivial profiles modify the vortex configuration
tending to increase its width. Increasing the vorticity, one is able to keep these effects under control.

  To conclude this Section, it is also interesting to notice that, in the case of small vorticity, the first derivative of  $\hat A_r$ does not
vanish at the origin for our solutions (see the  first three plots of Fig. \ref{sols}). This does not correspond to any  singularity for such field at $r\,=\,0$, 
though, 
 since the 
  equation of motion for $\hat A_r$, eq \eqref{sys8c}, is algebraic and exactly solvable along the  entire  
   radial direction. 
    We interpret the non-vanishing slope
     for the profile of  $\hat A_r$  at $r\,=\,0$ as supported by the modified 
     slope of the  real part of the  Higgs profile at the origin -- that
      is sourced by the presence of  $\hat A_r$, see eq \eqref{sys8a} -- in such a way to have a regular solution everywhere. The profile for  $\hat A_r$ does acquire
      a vanishing first derivative when increasing vorticity: see the last plot in Fig. \ref{sols}.
          
 \subsection{Anatomy   of the vortex }
\begin{figure}[!htb!]
\begin{center}
\includegraphics[width=0.23\textwidth]{nophase} \ \includegraphics[width=0.23\textwidth]{l6phase} 
\
\includegraphics[width=0.23\textwidth]{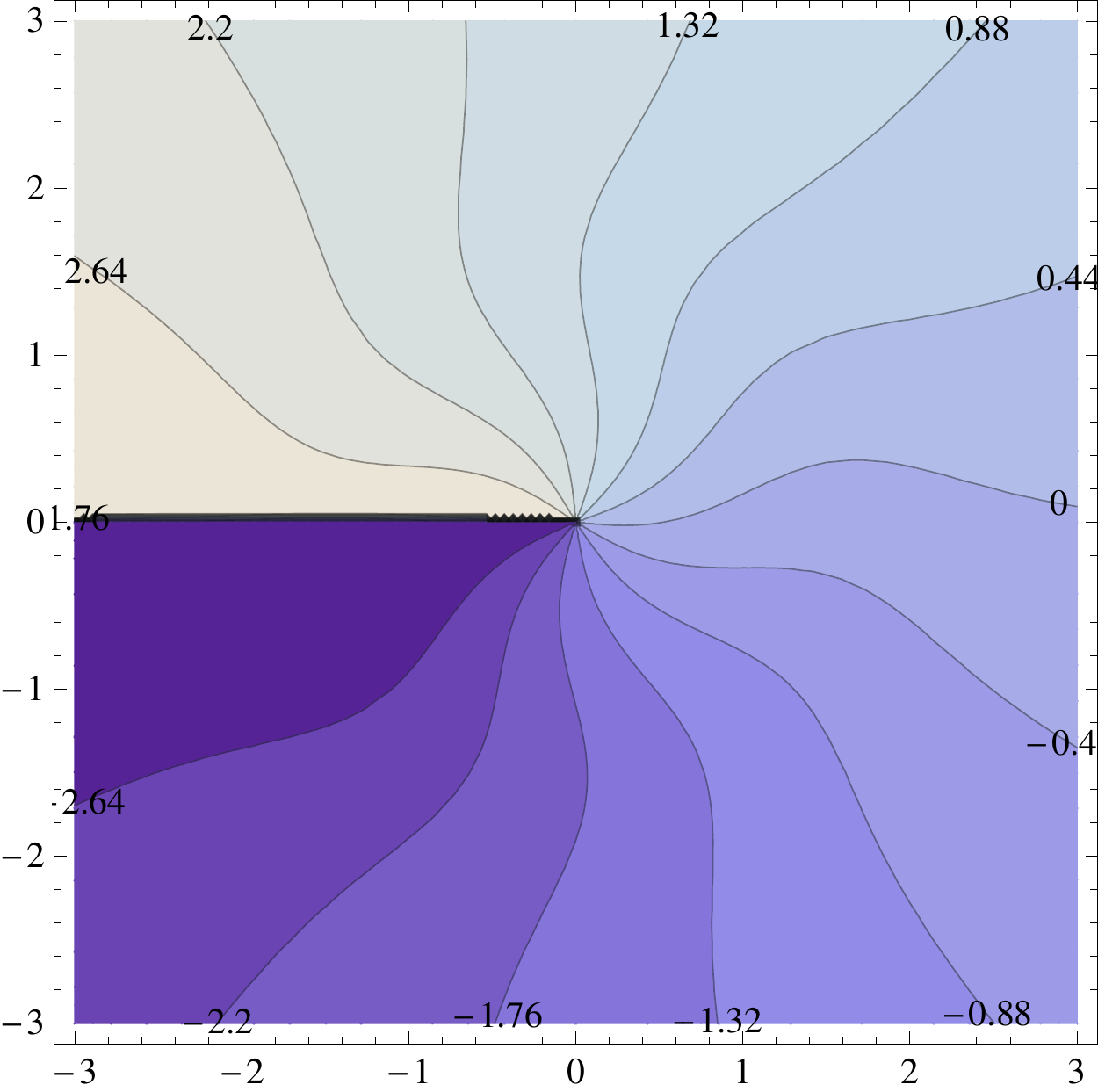}
\
\includegraphics[width=0.23\textwidth]{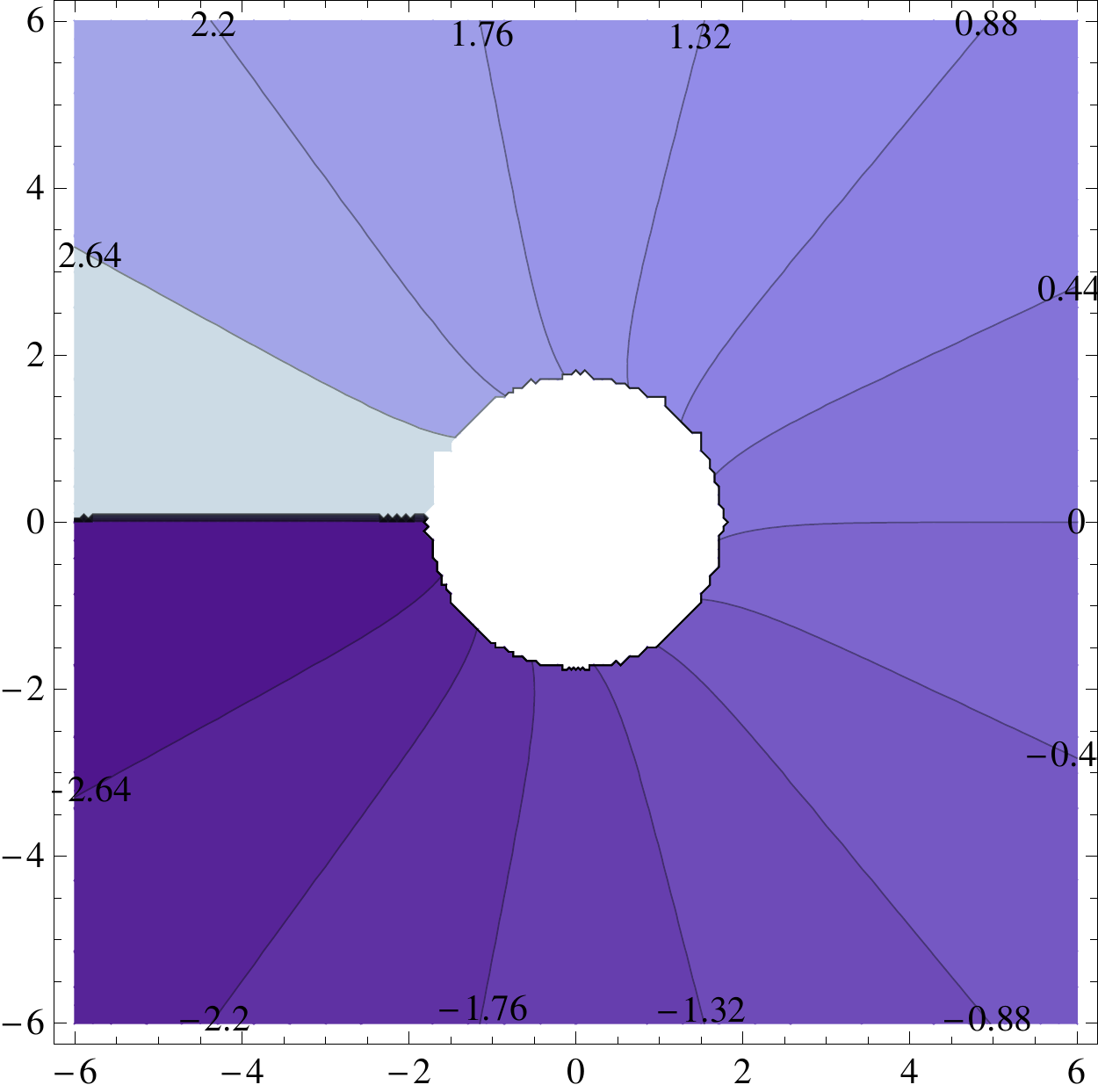}
\caption{\it{From left to right, lines of constant values of the Higgs phase for a) a NO vortex, 
b) a vortex computed in the limit of small $\beta$, where the backreaction of $\hat A_r$  onto
the other fields can be neglected, c) a numerical solution  with $\beta=0.40$ and $N=2$, and d) a numerical solution with $\beta=0.50$ displaying a `thick' singularity. For the last
two solutions we chose relatively large values of $\beta$ in order to make the effects of non-linearities more evident. Notice
how non-linear effects associated with our derivative interactions qualitatively change the Higgs phase profile.}}\label{sing-fig}
\end{center}
\end{figure}

 The distinguishing feature of our vortex configuration is the fact that gauge invariant field
 profiles  break  the reflection symmetry around an axis,
 accompanied by the complex conjugation of the scalar. This is a qualitatively new
 effect absent for NO configurations. 
{Given a vorticity  $N$,} 
  for  sufficiently large values of the parameter $\beta$ the solution becomes singular near the core, since the argument
  of the square root in eq. \eqref{rootm} becomes negative hence the solution becomes imaginary.  
   The vortex ceases to exist, and a `thick singularity' develops at the core of the configuration. 
{This limits the allowed vorticities for a given $\beta$.} We discuss  these properties by adopting a specific gauge that makes them easier to 
  study.  
 From its definition \eqref{ansatz1b},  we see that the gauge invariant quantity  $\hat A_r$
    is formed by combining two quantities that are gauge dependent:
    
    \be
    \hat A_r\,=\,e\,A_r+\partial_r \chi
   \,. \ee
    If one wishes to select  a particular gauge,
   a non-vanishing  $\hat A_r$ can be attributed to a non-vanishing radial component of 
    the gauge field, {\it or} to a radial dependent Higgs phase. Each one of the two cases 
   can be instructive, depending on the purpose. Here we focus on the case in which 
    the vector component vanishes,  $A_r=0$, while
   a radial dependent
    contribution to the phase $\chi\,=\,N\theta+\tilde{\chi}(r)$ 
    is turned on. We represent in Fig \ref{sing-fig} the scalar phase  for 
vortices with increasing
    values of $\beta$. The lines correspond to lines of constant phase:  $N \theta+\tilde{\chi}(r)\,=\,\text{constant}$. The first three plots represent regular solutions, the last one a singular
    configuration. 

It is clear that for Galileon Higgs vortices the figure is not symmetric under a reflection around the $x_1$ axis. At sufficiently large values
of the parameter $\beta$, the solution of the phase ceases to be well defined in the entire radial coordinate, and
 the vortex core is substituted by  a `thick'  singularity (recall that we graphically met this phenomenon also in Fig \ref{break} when discussing the profile for $\hat A_r$). 
 
 It is interesting to speculate what are the physical
 consequences of this fact. In particular  what  happens in the interior part of the thick singularity,
  that we define as the cylindrical surface with 
 boundary at $r=r_c$ where the square root turns complex. Possibly, a solution with the same Ansatz \eqref{ansatz3}
 as the one we considered arises, but with different `vorticity'  (in the sense that while the exterior solution has asymptotically a vorticity (say) $N=1$, the interior solution satisfies boundary conditions at the origin that correspond
 to higher vorticity $N>1$). Such
 configuration would be well-behaved for $r\to0$, and then would continuously connect with the exterior solution 
 at the core surface $r_c$. However, in trying to explicit determine
 the solution,  we  numerically find that some  of the field first derivatives are discontinuous at $r=r_c$. 
 
 { Hence in these theories the system seems to need a sort of `thick brane' regularization of a  singularity. It might be that such systems are related to -- and 
 find applications for -- the SLED proposal pushed forward by Burgess and collaborators (see e.g. \cite{Aghababaie:2003wz}), that makes use of the properties of codimension two object 
 for addressing the cosmological constant problem. We  leave these questions to further study.
}

\subsection{The energy functional for the galileon vortex}

It is a natural question to ask whether our configurations are stable under small perturbations. For the case of Abelian
Higgs vortex, the energy functional is known to be bounded from below: a BPS bound exists
corresponding to
a minimum for the energy. 
 For the case of Galileon vortex, a similar result does not hold: the existence of a BPS bound is not automatic, and
 additional assumptions on the configurations considered have to be imposed. On the other hand,  our configurations
 are characterized by non-vanishing vorticity -- a topologically conserved number -- hence they cannot continuosly change and decay to zero vorticity configurations. 
  Moreover,  we  are  able to show that the energy density is bounded for the static solutions we considered in the previous section.
 
 \smallskip

First, we  discuss the issue of a  BPS bound for a Galileon Higgs vortex.  
In section \ref{boundah} we saw that when the BPS bound, $\lambda = 2 e^2$, and the self-dual equations \eqref{sdeq} are satisfied,  the energy functional for the Abelian-Higgs Lagrangian
reaches the minimum 
$$ \mc E_{AH} = e \eta^2 \left| \int d^2 x B  \right|.$$
We now analyze  self-dual equations for the Abelian Higgs-Galileon Lagrangian \eqref{lahg}.  For this purpose, the first step is to write the Lagrangian for  Galileon Higgs in terms of the derivative operators $D_{\pm}\equiv D_1 \pm i D_2$, this gives
\begin{align}
\mathcal L_{6} &= \mathrm{Im}\left\{ -\frac{1}{4}\left( |D_+ \Phi|^2+|D_-\Phi|^2 \right)\Phi^\star D^2\Phi\right. \nonumber \\ &\left.- \frac{1}{4}\left[( D_{+}\Phi) (D_{-}\Phi)^\dag\Phi^\star D_{-}D_{-}\Phi +(D_{+}\Phi)^\dag (D_{-}\Phi) \Phi^\star D_{+} D_{+}\Phi \right]\right\}.
 \label{enfunl6}
\end{align}
In order to isolate the explicit dependence on the magnetic field, we use $D^2 = D_\mp D_\pm \mp i B e$, which can be proved by expanding $(D_1\mp iD_2)(D_1\pm iD_2)$. It also follows that  $2 D^2 = D_-D_+ + D_+ D_-$. Assuming that all the fields are static, the total energy functional is  the negative of the spatial part of the total Lagrangian \eqref{lahg},
  \begin{align}
\mc E_{AHG} &= \int d^2 x\left[\frac{1}{2}\left( B\mp e (|\Phi|^2 - \eta^2) \right)^2 + |D_{\pm}\Phi|^2 + \left(\frac{\lambda}{4} - \frac{e^2}{2} \right) \left( |\Phi|^2-\eta^2 \right)^2 \mp e \eta^2 B \right. \nonumber \\ 
&\left.  \mp \frac{\beta}{4}\left( |D_+\Phi|^2 + |D_-\Phi|^2 \right)|\Phi|^2 B e +\frac{\beta}{4} \left( |D_+\Phi|^2 + |D_-\Phi|^2 \right)\text{Im} [\Phi^\star D_\mp D_\pm \Phi ]+\frac{\beta}{4}M_{ab}   \right],\label{enfuncahg}
\end{align}
where we have defined
\begin{equation}
M_{ab} =\text{Im}\left[ ( D_{+}\Phi) (D_{-}\Phi)^\dag\Phi^\star D_{-}D_{-}\Phi +(D_{+}\Phi)^\dag (D_{-}\Phi) \Phi^\star D_{+} D_{+}\Phi \right] \, .
\end{equation}

The candidate for a  self-dual point of the AH and of the AHG models is the same: indeed the potential term of $\mc E_{AHG}$, which we identify by the coefficient $\lambda$,  
is cancelled  at $\lambda = 2 e^2$.  However, a relevant difference is that for the AHG
 model the energy functional at the self-dual point is not automatically bounded from below
 since the last two terms in  \eqref{enfuncahg} are not automatically positive definite. 
A sufficient (but by no means necessary) condition to have an
  energy functional  bounded from
below is
\begin{equation}
\frac{\beta}{4} \left( |D_+\Phi|^2 + |D_-\Phi|^2 \right)\text{Im} [\Phi^\star D_\mp D_\pm \Phi ]+\frac{\beta}{4}M_{ab} \geq c\,, \label{condc}
\end{equation}
for a certain constant $c$ which can be negative as long as the total energy remains positive. This is  the simplest
way we found to make sure that the Abelian Higgs-Galileon system has an energy 
 density bounded from below \textemdash other possibility might exist though. Given these preliminary results, it would be interesting
 to study more in general the stability of our configurations under small fluctuations, 
 and to explore alternative methods to obtain the BPS equations for Galileon
vortices, such as methods based on the energy-momentum tensor \cite{deVega:1976mi}
or on the Lagrangian of the system rather than on the Hamiltonian \cite{Atmaja:2014fha}.
 \begin{figure}
\begin{center}
\includegraphics[width=0.45\textwidth]{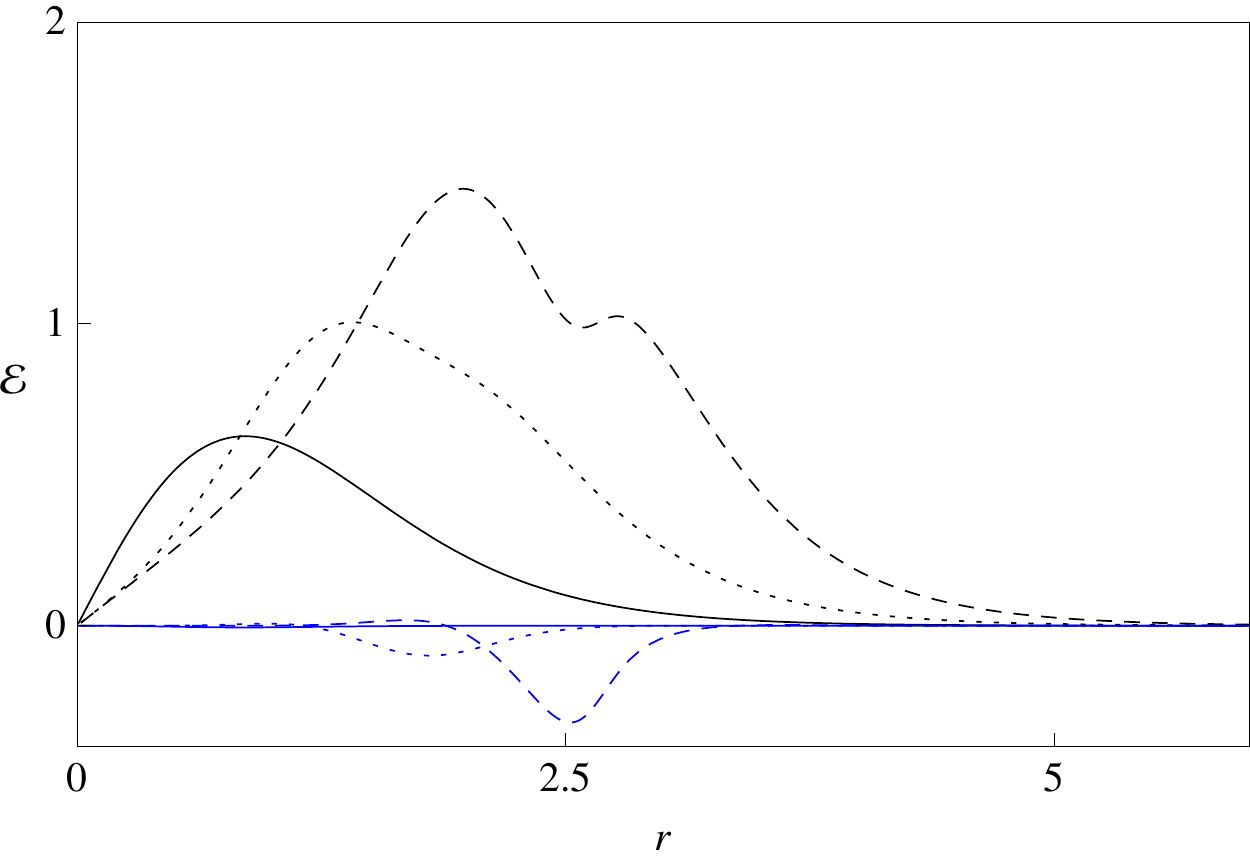}\caption{\it{The total energy functional for 
$(\beta=0.1, N=1), (\beta=0.4, N=2)$ and  $(\beta =0.5, N=3)$, is shown in solid, dotted and
dashed lines respectively. The energy functional always develops a minimum
at $r=0$. As $\beta$ increases, the energy density concentrates farther away from
the origin and the total energy increases. In blue, we show the contribution from $\mathcal L_6$.}}\label{energy}
\end{center}
\end{figure}

\smallskip

We end this section showing that 
 the static  configurations discussed in section \ref{exsol} have positive definite
energy for the values of $\beta$ we considered. 
%
We do not use the Lagrangian in terms of $D_\pm$,  eq. \eqref{enfuncahg}, but rather we work directly with
the static axially-symmetric ansatz. Then 
\begin{equation}
\mc E_{AHG}[\hat A_a(r), X(r)] = -2\pi \int_{0}^{\infty} \left (\mc L_{AH}[\hat A_a(r), X(r)] +  \mc L_{(6)}[\hat A_a(r), X(r)] \right) r dr .
\end{equation} 
The integrand of this expression is plotted in Fig. \ref{energy}.  The contribution to the energy coming from $\mathcal L_{(6)}$ is negligible only for $\beta=0.1$, and it is centred around the
region where the non-linearities are relatively large, creating a local minimum.
For any $\beta$ the energy of the vortex is finite and it always develops a global minimum 
at the locus of the vortex. If the non-linearities due to the derivative couplings are too large, 
we can suspect that the local
 minimum due to $\mathcal L_6$  {approaches to zero, and if it tries to drive the total 
energy below zero then a thick singularity is formed. There might be well-behaved complete
 solutions where there are two global minima, both of them at $\mc E_{AHG}=0$, however we could not find numerically such solutions.   }

\section{Coupling with gravity}

It is known that a NO vortex coupled to gravity backreacts on the geometry, generating
  a  space-time  with a  
 conical singularity  when seen by a distant observer (see e.g. \cite{Gregory:1987gh,Gregory:1997wk}).
   In this section we study the coupling to gravity of a Galileon Higgs vortex. {  We are 
    interested to determine the gravitational backreaction 
    of the field profiles of the vortex configuration that we determined in the previous sections. 
    We will learn 
     that, in a sense, Einstein equations `suggest' a field dependent change of coordinates adapted to the vortex profile, that  makes the resulting geometry particularly symmetric \footnote{
     We thank Ruth Gregory for useful remarks on the content of this section.}. }

   For our  purpose,  we consider the Einstein equations minimally coupled to the
energy momentum tensors of the Abelian Higgs model and of the  Higgs Galileon contributions. Despite we work in four spacetime dimensions we consider only the Galileons given by $\mc L^{(6)}$. 
 In this way, we avoid the issue of having to include non-minimal couplings with gravity, that
 would be necessary to maintain a ghost-free condition \cite{Hull2,Tasinato:2014eka,Heisenberg:2014rta}.
 %

The energy momentum tensor for the AH model is
\begin{align}
T^{(AH)}_{{a}{b}}& = \frac{2}{\sqrt{-g}}\frac{\delta (\sqrt{-g} \mc L_{AH})}{\delta g^{{a}{b}}} = -g_{{a}{b}} \mc L_{AH} +2 \frac{\delta \mc L_{AH}}{\delta g^{{a}{b}}} \nn \\
& = -g_{{a}{b}} \mc L_{AH} + 2 \eta^2\covD_{a} X \covD_{b} X + 2 \eta^2 X^2 \hat A_{a}
\hat A_{b} + \frac{1}{e^2} F_{{a}c} F^c{}_{b},
\end{align}
while for the  Galileon Higgs we obtain
\begin{equation}
T^{(6)}_{{a}{b}} = -g_{{a}{b}} \mc L_6 +2 \beta \eta^4 \left( L_{{a}{b}} Q + L Q_{{a}{b}}   - Q_{{a}}{}^c{} L_{c{b}} - L_{{a}c}Q^{c}{}_{{b}} \right)  .
\end{equation}

We restrict ourselves to small $\beta$ coupling, and to  weak coupling to gravity.  
In particular we are interested in establishing  how the spacetime metric  can take  into account the
breaking of reflection symmetry shown by the Galileon Higgs vortex profiles. Inspection of the energy momentum tensor reveals that -- for field configurations
 corresponding to a vortex -- the profile for  the gauge invariant  field  $\hat A_r$ induces a component $T^{(AH)}_{r\th}$, which is not supported by the Einstein tensor relative to 
a diagonal metric. For this reason, we take a metric Ansatz of the form
\begin{equation} \label{metan1}
ds^2=e^{2(\gamma-\Psi)}(dt^2-dr^2)-e^{2\Psi}dz^2 - \alpha^2 e^{-2\Psi}d\th^2 - \beta\,  \omega dr d\th,
\end{equation} 
where $\gamma$, $\Psi$, $\omega$ and $\alpha$ are functions only of $r$. The parameter $\beta$ -- the same that multiplies $\mc L_6$ -- is taken 
to be small.   { As we will
see in a moment, 
such form of the metric, breaking the reflection symmetry $\theta\to- \theta$,  is in principle able to accommodate for the specific field profiles we are considering.  On the other hand,
this metric is still axially symmetric, since all the metric components depend only on the radial direction. As we will discuss towards the end of this section, a coordinate transformation
exists that is adapted to the vortex  configuration, and that renders \eqref{metan1} explicitly diagonal.}

 \subsection{Solving Einstein equations in a small $\beta$ limit}

For the moment we work with the metric \eqref{metan1}, to investigate how the non-diagonal metric component $\omega$ depends
on the field profile. 
 The Einstein equations controlling  gravity minimally coupled to the vortex are
\begin{equation}
G_{ab}=8\pi G(T^{(AH)}_{ab}+T^{(6)}_{ab}).
\end{equation}
Boost invariance along the $z$-axis is automatically satisfied for the energy-momentum tensor of the Abelian Higgs model (i.e. $T^{AH}{}^{t}\!{}_{t}=T^{AH}{}^{z}\!{}_{z}$), however it only holds for the  Galileon Higgs if we impose the condition
$\gamma = 2\Psi$, which we will do from now on.  
As we explained, we take a small $\beta$ limit: the configuration of $\hat A_r$  can be easily obtained from eq \eqref{rootm} by expanding at first order in $\beta$:
\be \label{rootm1}
\hat A_r \simeq
\left(\frac{6 \beta \eta^2 X^2}{r}\right)\,
\left[
\frac{\hat A_\theta}{r^2}( \hat A_\theta -  r    \hat A_\theta'{})+\frac{X'}{3X }\left( \frac{ X'{}}{X} - \frac{4 \hat A_\theta^2  }{r} \right)\right]
%
%
\ee
and is then proportional to the quantity $\beta \eta^2$.  In addition, in this limit
$X$ and $\ha_\th$ describe standard NO vortices. 

At our level of approximation -- leading order in $\beta$ -- 
 the non-vanishing components of   the total energy momentum tensor are only in the Abelian Higgs sector, and 
 are given by:
\bse
\begin{align}
T^{AH}_{tt} &=\mc P_t  =  \frac{1}{4}e^{\gamma}\lambda\eta^4(X^2-1)^2 + \frac{e^{2\gamma} \eta^2 \ha_\th^2 X^2}{\al^2} + \frac{e^{\gamma}\ha_\th'{}^2}{2 q^2 \al^2} +  \eta^2 X'{}^2, \\
T^{AH}_{rr} &=\mc P_r   = -\frac{1}{4}e^{\gamma}\lambda\eta^4(X^2-1)^2 - \frac{e^{2\gamma} \eta^2 \ha_\th^2 X^2}{\al^2} + \frac{e^{\gamma}\ha_\th'{}^2}{2 q^2 \al^2} + \eta^2 X'{}^2, \\
T^{AH}_{\th\th}&=\mc P_\th   = - \frac{1}{4}e^{-\gamma}\lambda\eta^4\al^2(X^2-1)^2 + \eta^2 \ha_\th^2 X^2 + \frac{e^{-\gamma}\ha_\th'{}^2}{2 q^2} - e^{-2\gamma} \eta^2 \al^2 X'{}^2, \\
T^{AH}_{zz} &=-\mc P_t   =  -T^{AH}_{tt},\\
T^{AH}_{r \th}&=\beta \mc M   =\beta \frac{\eta^2\omega}{2}\left[ \frac{4 \eta^2 \ha_r^{(0)} \ha_\theta X^2}{\omega} -\frac{1}{4}\eta^2\la (X^2-1)^2 -\frac{e^\gamma  \ha_\theta^2 X^2  }{ \al^2} +\frac{ \ha_\th'{}^2}{2 q^2 \al^2\eta^2} - e^{-\gamma}  X'{}^2       \right].\label{eq:m}
\end{align}
\ese
where we have defined  $\hat A_r\equiv\beta\eta^2 \hat A_r^{(0)}$. We emphasize
that the fields $X$ and $\ha_\th$ have profiles corresponding to  a NO vortex, since we are neglecting the ${\cal O}(\beta^2)$ effects associated
with the backreaction of $\hat A_r$ on their equations of motion. 

Thus, the only consequence of the presence of $\mathcal L^{(6)}$ is that for Galileon Higgs vortices $\ha_r$ is necessarily different from  zero and therefore $T^{AH}_{r \th}$ cannot be turned off.  It is convenient to rescale $r\to\eta^{-1} r $ and $\al\to\eta^{-1}\al$, so that $\eta$ effectively
controls the coupling strength between the vortex and the Einstein tensor,
\begin{align}
\al''&=-4\pi G \eta^2 \al (\mc P_0-\mc Pr), \nn \\
(\al\gamma')'&=4\pi G \eta^2 \al(\mc P_r + \frac{e^{2\gamma}}{\al^2}\mc P_\th),\nn\\
\al'\gamma'&=\frac{ {\al}\gamma'{}^2}{4}+4\pi G \eta^2 {\al} \mc P_r,  \nn \\
\frac{1}{8}e^{-\gamma} \omega (\gamma'{}^2 + 4 \gamma'')& = 4 \pi G \eta^2 \mc M .\label{eqs:einstein}
\end{align}
Note that $\omega$ is completely determined by the solutions for the other fields,  and it 
vanishes when $\ha_r^{(0)}=0$ because its equation of motion acquires the form
$$\omega F(X, \ha_\theta,\al,\gamma, X', \ha_\th',\gamma')=0,$$
where the function $F$ of the various  fields  does not generically vanish  for vortex configurations.   
To the lowest order in $\eta^2$ and in the metric corrections, we find from the first, third, and fourth equations in \eqref{eqs:einstein} the solutions
\bse
\begin{align}
\al & = \left[1-4 \pi G \eta^2 \int r (\mc P_0 - \mc P_r) dr \right] r + 4 \pi G \eta^2 \int r^2 (\mc P_0 - \mc P_r) dr,  \\
\gamma & = 4 \pi G \eta^2 \int r \mc P_r dr,  \\
\omega & = \frac{2 \pi \mc M }{(r \mc P_r)'}. \label{eqom}
\end{align}
\ese
Using the Bianchi identity $r^2 ( r \mc P_r)' = \mc P_\th + \mc O(\beta^2)$ we can verify that  the second equation in \eqref{eqs:einstein} is also satisfied by these solutions, and we can rewrite the solution for $\omega$ as
 $\omega\sim r^2 \mc M/\mc P_\th$.

The fields in the Galileon Higgs vortex decay fast, therefore the integrals in the solutions for the metric components quickly reach their asymptotic, constant values. To verify that
$\omega$ is well behaved asymptotically we can consider the expression for $\hat A_r$ in the small $\beta$ limit, eq \eqref{rootm1}.
Since $X(r\to\infty)\sim 1$ we see that $\ha_r$ decays at least as $\ha_\theta/r^3$. 
 Using this
information we learn  that the first term in \eqref{eq:m} is sub-dominant with respect
to the other terms, so that  $\mc M/\mc P_\theta$ decays as $ \omega/r^{2}$, 
and the only solution to eq. \eqref{eqom} is $\omega = 0$.
 This implies that for large $r$ the metric has the same form as the metric that would be obtained in the
 presence of a weakly coupled NO vortex, which was derived in \cite{Gregory:1987gh} and
corresponds to a conical metric with a deficit angle $\Delta = 8 \pi G \mu$ as seen by an asymptotic observer, where  $\mu$ is the energy per unit length of the string.

With a little additional effort, we can also derive the asymptotic profiles for the fields
involved, within our  approximations. 
 We start 
 considering the asymptotic solutions for the profiles of $X$ 
 and $\hat A_\theta  $ for the   NO vortex, also valid for our configuration, at leading
 order in a small $\beta$ expansion    (see e.g. \cite{Rubakov:2002fi}):
\begin{equation}
X\approx 1 - x_0 \frac{e^{- \eta \sqrt{\la} r}}{\sqrt{r}}, \ \ \ 
\hat A_\theta \approx a_0 \sqrt{r} e^{-\sqrt{2} e \eta r},
\end{equation}
where $x_0$, $a_0$, and $b_0$ (used in the next equation) are constants determined by the boundary conditions.

Plugging these solutions in \eqref{rootm1} we get
\be
\hat A_r \approx \frac{b_0}{r}{e^{-2 \sqrt{2} e \eta r}},
\ee
Using these results and the background metric to evaluate $\mc M$ and $\mc P_\theta$
we find that asymptotically $\omega$ is given by
\begin{equation}
\omega \approx  -\frac{1}{r^{3/2}} e^{-2 \sqrt{2} e r \eta }.
\end{equation}
Hence we see that it has an exponential decay for large values of $r$.

\subsection{A convenient coordinate transformation} 
 {So far, working
  at linearised order in $\beta$, we have shown that the field profile for $\hat A_r$ turn on a new metric component when coupling with gravity, that
  we denote with $\omega$ in eq \eqref{metan1}.
  For 
 concluding this Section, we show that taking advantage of the invariance under diffeomorphisms
of General Relativity we can perform a
  change of coordinates
  that renders the
  metric diagonal and  manifestly 
   axially symmetric.\footnote{
    Let us point out that, in absence of 
   gravity,  it is also possible to make a    
choice of  coordinates  that removes the radial component of $\hat A_a$. However, the resulting flat 
 space-time would correspond to Minkowski space expressed 
 in a very convoluted 
  coordinate system, that would render more complicated  the analysis of the properties of our configuration, and the comparison with the NO vortex. 
}
   Einstein equations  relate the magnitude of the metric component $\omega$ -- that breaks the reflection symmetry $\theta\to-\theta$ in metric \eqref{metan1} -- 
    with the size of the field $\hat A_r$, that as we learned is producing the twirling features of the vortex configurations.  On the other hand,
always working at leading order in $\beta$,    the  following redefinition  of the angular coordinate
    \be
 {d   \theta\,\to\, d \theta-\frac{\beta\,\omega}{2\,\alpha^2}\,e^{2\psi} \,d r}
    \ee
    renders  the geometry  manifestly axially symmetric, giving it a  diagonal form. Such field redefinition adapts the geometry to the vortex configuration, and effectively `eats up' the contribution
    of the field $\hat A_r$ that would cause an off-diagonal component $T_{r\theta}$ in the energy momentum tensor.   
    Hence in this specific coordinate system, the  coordinates adapt  to the vortex
   lines, 
     and the derivative interactions modulate the radial dependence of $\hat A_\theta$, $X$.  
     It would be interesting to investigate whether the arguments we developed in this section can be extended to arbitrary values of $\beta$, to understand  the 
       gravitational backreaction in large $\beta$ regimes.}

\section{Outlook}

In this work we presented and analysed  finite energy vortex solutions in a 2+1 dimensional  Abelian Higgs model supplemented by higher
order derivative self-interactions for the Higgs field. 
 Such interactions have been first introduced in \cite{Hull:2014bga} for providing a Higgs
mechanism to spontaneously break the gauge symmetry through  vector Galileon interactions \cite{Tasinato:2014eka,Heisenberg:2014rta}.  They are ghost free, and in a suitable high energy limit they enjoy Galilean
symmetries that can help for protect their structure from quantum corrections. We dubbed this system Galileon Higgs.  
Within this framework, we have been able to determine vortex solutions characterised by topologically conserved
winding numbers. They have features that make them qualitatively different from the Nielsen-Olesen vortex. 
The derivative  non-linear interactions turn on  new   field profiles for gauge invariant field potentials  that   violate a reflection symmetry around one
of the axis of the Cartesian coordinates,  and lead  to regular configurations
that necessarily  also break the axial symmetry of the configuration.
Interestingly, some of the equations of motion 
reduce to quadratic algebraic equations,  simplifying  considerably our analysis. 
Moreover, we have also promoted our 2+1 dimensional solution to a 3+1 dimensional  one, and coupled the resulting system
to gravity,  showing  that gravity backreaction leads to
 {a  space-time that, depending on the coordinates one choose, can be 
described by a metric without reflection invariance, or a metric with non-standard
angular and radial coordinates. One way or another, the effect of having a vortex with
 non-trivial field profiles is seen as a contribution to the space-time curvature
and  deficit angle.}

\smallskip

Our results can find several applications and suggest further lines of research,  opening possibilities for finding new classes of vortex solutions
in system with derivative self-interactions.
 For example:
 \begin{itemize}

\item[$\blacktriangleright$]  It would be interesting to find non-relativistic analogues of our
 Galileon Higgs vortex configurations, for example in the context of superfluids or
 superconductors. Such non-standard
 vortex configurations might play some role in cases in which derivative interactions are important
 in the pattern of symmetry breaking. Also, in this context, the dynamics of multi-vortex solutions would be interesting 
 to investigate, since it can be important when discussing the stability of our configurations when considering  values 
 of the vorticity larger than one.

\item[$\blacktriangleright$] 
 As discussed in the introduction, one  motivation for
studying cosmic strings/vortex solutions in this context  is to understand 
screening mechanisms -- as Vainshtein mechanism -- in absence of  spherical symmetry, taking
into account the backreaction of all fields including gravity.  This  can be important  when testing screening mechanisms in the context
of cosmology as for understanding the cosmic web  structure, where filaments and voids form (see e.g. \cite{Falck:2014jwa,Falck:2015rsa}). 
{ Our results  suggest that in some cases -- depending on the field content and their interactions -- the gravitational 
  backreaction can be rather subtle, and 
  axial symmetry of the system can be  not manifest even for  cylindrical sources.
   These findings 
   might offer indications for determining accurate  semi-analytical models for structure formation in models
 with screening
 mechanisms.  }

 \end{itemize}
 We plan to develop
  these arguments in further studies. 
  
 \acknowledgments
We are happy to  thank  Ruth Gregory,   Gustavo Niz, Ivonne Zavala, and an anonymous referee for useful discussions or comments on the draft.  JC thanks support by the grant CONACYT/290649. 
GT is supported by an STFC Advanced Fellowship ST/H005498/1. He thanks the University of Leon Guanajuato
for kind hospitality during the course of this work, and the Mexican Academy of Science for generous financial support.

\end{document}